\documentclass[11pt,titlepage]{estech}

\usepackage{amsmath}
\usepackage{enumitem}
\setlist[description]{leftmargin=2\parindent,labelindent=\parindent}
\usepackage{graphicx}
\usepackage[numbers]{natbib}
\usepackage{subcaption}
\usepackage{textcomp}
\PassOptionsToPackage{hyphens}{url}
\usepackage{hyperref}  
\newcommand{\improveFigureLayout}{
\renewcommand{\topfraction}{0.85}
\renewcommand{\textfraction}{0.1}
\renewcommand{\floatpagefraction}{0.75}
\renewcommand{\dbltopfraction}{\topfraction}
\renewcommand{\dblfloatpagefraction}{\floatpagefraction}
}

\newcommand{\code}[1]{\texttt{#1}}
\newcommand{\ipaddr}[1]{\texttt{#1}}
\newcommand{\term}[1]{\textit{#1}}

\newcommand{\Threatcases}{Cases}

\newcommand{\threatcasesTM}{ThreatCase\textsuperscript{\tiny\textregistered} results}
\newcommand{\threatcase}{case}
\newcommand{\threatcases}{cases}
\newenvironment{interview}{\begin{list}{*}{\setlength{\labelwidth}{1cm}%
  \setlength{\leftmargin}{1.6cm}
  \setlength{\labelsep}{0.5cm}
  \setlength{\rightmargin}{1cm}
  \setlength{\parsep}{1ex plus0.5ex minus 0.1ex}
  \setlength{\itemsep}{0.5ex plus0.2ex}}}
  {\end{list}}
\newcommand{\interviewer}{\item[\textbf{I:}]}
\newcommand{\operator}{\item[\textbf{O:}]}
\newcommand{\question}[1]{\textsl{#1}}
\newcommand{\AppName}{esINSIDER}
\newcommand{\AppAbbrv}{esINSIDER}

\improveFigureLayout

\title{Reframing Threat Detection: Inside \AppName}
\shorttitle{Inside \AppAbbrv}
\author{
  M. Arthur Munson$^*$\thanks{art.munson@esentire.com},
  Jason Kichen$^*$,
  Dustin Hillard$^*$,
  Ashley Fidler$^*$,
  Peiter Zatko$^\dag$\\
  \\
  {\normalsize\textit{$^*$eSentire Inc., Cambridge, ON, Canada}}\\
  {\normalsize\textit{$^\dag$Advisor}}
}

\reportID{ES2019-001}
\date{Revision 1.0, \today}

\frontsummary{}

\begin{document}
\maketitle
\tableofcontents
\newpage

\section{Introduction}

Organizations are spending millions of dollars a year on computer and
network security products, yet we continue to see growing numbers of
large-scale compromises by persistent threats and insider
threats.\footnote{Aspects of \AppName\ are covered by claims of a pending
  patent application.}
When an industry throws more resources at a problem, but the situation
continues to worsen, it behooves us to step back and understand what
is missing.  Ignoring this new risk used to be a rational business
decision because only governments and large corporations needed to
worry about being targeted by persistent threats and insider threats.
Today, they are tangible risks for practically every business.

Current security approaches struggle to detect ongoing threat
campaigns because 1) they focus on individual events and alerts, 2)
they tend to overemphasize preventing (or detecting) initial access
and misuse, and 3) significant data storage, processing, and analytics
are required to aggregate and correlate events over weeks.  When
security analysts review individual alerts, they are reviewing
isolated information tidbits; it is exceedingly hard to see if they
fit into a larger story of a serious threat.  As a result, serious
threats are lost in the stream of false positives and low priority
problems~\citep{Schwartz14:TargetIgnored}.  The intense focus on
stopping initial access is problematic because the threat actor only
needs to succeed once, and they get an unlimited number of tries.  A
well-resourced threat actor will eventually get in; if defenders
ignore internal network patterns, the threat actor can operate inside
the network---the bulk of their campaign---with impunity.  In our
experience, only a minority of organizations watch for suspicious
internal network actions, and virtually nobody connects events and
alerts together over weeks to detect an ongoing campaign. As a result,
network defenders lack the information to form a clear picture of the
long-term threats unfolding in their networks.

To better defend against persistent and insider threats, defenders
need to start thinking like a threat actor to understand what they
must do to complete their campaign.  The threat actor, whether an
external actor or a malicious insider, has a goal of identifying,
collecting, and stealing data of value to their mission. The specific
techniques used to gain initial access and to maintain a persistent
foothold within a network change based upon a number of
variables. Hunting for evidence of tools and tactics that change so
frequently is doomed to be a game of catch up
(e.g.,~\citep{Dunwoody16:NoEasyBreach}).\footnote{There are times when
  open source or known implants are employed, when a 2-year old CVE
  will suffice, and proxying through already suspect bulletproof
  hosting providers is appropriate. No more or less common are the
  times when custom, bespoke tooling is required, when the exploit
  chain includes more than one 0day, and when the infrastructure is
  acquired just-in-time, specific to the operation at hand, and
  designed to blend into the noise of the Internet. Equally common is
  the use of applications and tools that are part of the default
  operating system installation. Threat actors ranging from the least
  to the most sophisticated live somewhere along this spectrum of
  sophistication.} In contrast, the adversary campaign behaviors
inside a compromised organization, abstracted from the specifics of
the techniques used, remain consistent across campaigns and across a
wide range of threat actors.  We can reduce the detection challenge by
focusing on the required meta-goals of the threat actor and how they
differ from data flows and movements in legitimate business
operations.  Regardless of the actor’s specific choices---how to move
laterally, which exploit to throw, which credential to use, etc.---the
majority of their campaign focuses on identifying, collecting, and
extracting as much information of value (to the threat actor) from the
target organization as possible.  These campaign behaviors---internal
reconnaissance, collection, and exfiltration---are required and occur
repeatedly, creating multiple opportunities for the threat actor to
be detected.

In this paper we describe \AppName, an
automated tool that detects potential adversary campaigns.  \AppAbbrv\
aggregates clues from log data, over extended time periods, and
proposes a small number of \threatcasesTM\ for human experts to review.
The core ideas are to 1) detect fundamental campaign behaviors by
following data movements over extended time periods, 2) link together
behaviors associated with different meta-goals, and 3) use machine
learning to understand what activities are expected and consistent for
each individual network.  We call this approach campaign analytics
because it focuses on the threat actor's campaign goals and the
intrinsic steps to achieve them.  \Threatcases\ package together
related information so the analyst can see a bigger picture of what is
happening, and their evidence includes internal network
activity resembling reconnaissance and data collection.  Linking
different campaign behaviors (internal reconnaissance, collection,
exfiltration) reduces false positives from business-as-usual
activities and creates opportunities to detect threats before a large
exfiltration occurs. Machine learning makes it practical to deploy
this approach by reducing the amount of tuning needed.

It is important to understand how this approach fits into a complete
cybersecurity portfolio.  We can organize cybersecurity into four
broad strategies: 1) prevention, 2) rapid intrusion detection and
response, 3) threat hunting, and 4) proactive closing of gaps (not
discussed in this paper).

\term{Prevention} is all about preventing problems from starting, and it
includes all the best practices learned over decades: end-point
defenses, strong passwords, appropriately configured firewalls, least
privilege, two factor authentication, user training, reducing the
organization's threat surface, segmented networks, etc.  Organizations
obviously should invest in prevention: it stops many problems before
they become problems.  At the same time, prevention will never be
100\% effective because network defenders cannot achieve perfection.
The threat actor only needs to find one gap, only needs to find one
user who clicks on a phishing link, only needs to get lucky once.

\term{Rapid intrusion detection and response} acknowledges prevention will
sometimes fail and then tries to detect intrusions after they have
started.  It includes network intrusion detection (e.g., Snort, Bro),
endpoint detection, and log analysis.  Naturally, one wants to detect
intrusions as fast as possible, before a small problem becomes a large
problem.  Intrusion detection raises the bar and improves a network's
security, but the game is still asymmetric.  A sophisticated threat
actor with enough time and resources will find a way to avoid
intrusion detection systems.

\term{Threat hunting} proactively searches for undetected threats that are
already in the network and have avoided detection by front line
security tools.  \AppAbbrv\ is an automated tool to assist with the discovery
of these threats.  Unlike user entity behavior analysis (UEBA), \AppAbbrv\
uses campaign analytics.  This perspective is a paradigm shift in
detecting persistent and insider threats:
\begin{itemize}
  \item These campaign steps are almost always required for the campaign to
    succeed.  Skipping a step, to avoid detection, makes it harder for
    the threat actor to complete the mission.
  \item Campaign detection can focus on many fewer activities: just
    the ones that would directly execute campaign steps.  In a
    nutshell, we need to follow the data.  It almost does not matter
    if the threat actor invents a new technique; they still need to
    find, access, and attempt to exfiltrate the data.
  \item It inverts the asymmetry.  The threat actor needs to execute
    their whole campaign.
    Defenders only need to identify enough of a subset of the campaign
    to catch them.
\end{itemize}
As a result, defenders can have the advantage and stop playing catch up.

All of these approaches contribute to the same end goal: protect the
organization's digital and physical assets.
They are also complementary.  E.g., discoveries from campaign
analytics can improve prevention and rapid intrustion detection \&
response.  You should use all of them, and they are all on-going,
parallel programs.

The rest of the paper is organized as follows.  First, we present a
hypothetical interview with a sophisticated threat actor to help
explain the threat actor’s perspective and how they operate
(Section~\ref{sec:interview}).  Next, we describe the fundamental
requirements for effective detection of sophisticated threat actors
(Section~\ref{sec:requirements}).  Section~\ref{sec:approach} explains
the basic design of \AppAbbrv, and Section~\ref{sec:conclusion} concludes.
For information about the automated machine learning used in \AppAbbrv, see
Appendix~\ref{sec:autolearn}.

\section{Hypothetical Interview with an Operator}
\label{sec:interview}

The \textbf{Operator} has executed campaigns for state actors and
various other entities. She has been a member of and managed the teams
responsible for executing these campaigns, in addition to directing
other teams of engineers and developers responsible for the tools and
infrastructure required to execute threat actor campaigns.  She is
also experienced with operations based on human access, i.e. insider
threat.

\begin{interview}
  \interviewer \question{There is a ton of activity in the security
    industry.  E.g., every year, thousands of CVEs are
    disclosed\footnote{\url{https://www.cvedetails.com/browse-by-date.php}},
    and millions of new malware are
    found\footnote{\url{https://www.av-test.org/en/statistics/malware/}}.
    It seems like you need to move fast in security to keep up with
    the latest developments.  How much does this constant flurry of
    activity disrupt your operations?}

  \operator This may be surprising, but not very much, which is good
  news for me and my teams. The focus on initial access, signatures,
  etc. misses the majority of where I spend my time, focus, and energy
  during operations. The perimeter is important, and high
  walls with barbed wire on them is something that I have to
  defeat. But I really only have to defeat it once, and I get,
  essentially, an unlimited number of attempts to do so. By focusing
  on signatures and CVEs, the defender is overlooking the majority of
  where I spend my time and effort. They essentially overlook my
  campaign.

  \interviewer \question{Can you clarify what you mean by ``campaign''?}

  \operator The campaign is maintaining covert and/or clandestine
  access to a targeted network and repeatedly finding information of
  value over time. This isn't a singular act or one-time event;
  instead, I, or my team, spend time finding, collecting,
  exfiltrating, evaluating the exfiltrated data, and then possibly
  modifying or re-tasking the mission based on what was learned from
  the information we obtained.

  \interviewer \question{Can you give an example?}

  \operator Think about the OPM
  ``hack''\footnote{\url{https://bit.ly/2EEo9I5}  The OPM hack explained: Bad security practices meet China's Captain America.},
  the large healthcare and medical compromises, the ``Panama
  Papers''\footnote{\url{https://en.wikipedia.org/wiki/Panama_Papers}},
  or even how an insider may identify, collect, and abscond with
  sensitive documents when they leave to go work for a
  competitor. Everyone talks about what was lost and the way I got in,
  but nobody talks about the main effort of my job which is what I was
  doing for days, months, sometimes years, while I'm inside.

  The initial access is just the first of many steps, several of which
  can happen simultaneously. I'm looking for things like running
  services and open ports, in order to figure out what kind of
  hardware and software is deployed on the network (since this then
  largely dictates the types of tools and tradecraft I will use). This
  internal network reconnaissance helps illuminate the network for
  me. In particular, I want to know where the different repositories
  of data are and how this target information is accessed within the
  environment. It's all about the data that I'm looking for. Once I
  start to locate and collect data that I want to get out, I need to
  move that data from wherever I found it to somewhere from which I
  can safely exfiltrate it. Because look, I'd love to simply throw the
  data\footnote{Ed: A throw is a transmission of a sizeable amount of
    perceived valuable data from inside the compromised network to an
    external destination of the threat-actors choosing.} from where it
  lives internally to my server externally, but Murphy is always
  around.  Inevitably, I can't just exfiltrate it from where I found
  it---or at least not without drawing a big bullseye on myself. So
  now I'm moving sometimes small, sometimes large volumes of data
  around the network at a particular point in time. However, over the
  duration of my operation this is almost always a large amount of
  data. And finally, I'll exfiltrate this data out to a system I
  control. The key, though, is that this isn't a serial process, or a
  process that occurs only once. It happens multiple times, sometimes
  in various orders, and because nobody is looking for me doing this,
  I do it for as long as I can, which sometimes goes on for years.

  \interviewer \question{OK, what does a campaign look like from your
    perspective?}

  \operator There can be a lot of complexity in it, to be honest. From
  start to finish, there is sometimes enormous technical and
  tradecraft issues to be discussed and decided. Networks are so
  dynamic day over day that my own tool and tradecraft choices aren't
  even remotely static, which works to my advantage if defenders are
  just looking for a particular tool’s signature.

  But looking at everything I and/or my team does, you can really
  bucket it all into some key areas. First, let's ignore the initial
  intrusion because---let's be honest---eventually I get in.

  So, once I'm in, the first thing to consider is internal network
  reconnaissance [stage 3]. As I said before, this is where we're
  discovering things about the network in order to figure out where
  the different repositories of data are and how we will need to
  access them.

  As we start to find data that we want to get out, we start moving it
  around so that we can eventually get it out. This internal data
  movement [stage 4] is where I am probably the most exposed but
  also the least concerned, since no one really watches the east-west
  data flow very closely. Also, at this point, I'm using legitimate
  credentials that I've obtained to get the data. If the defenders
  looked for one of their own systems hoovering up information from a
  range of internal servers, rather than trying to figure out when or
  how I stole Joe-the-admin's credentials, I'd have a really hard
  time.

  At some point I'll start exfiltrating data [stage 5], both the
  ``operational intelligence'' sort of data (low volumes, but things
  like screenshots, keylogs, etc.), and the big fish collections like
  the massive, sensitive database dumps. This is where some people are
  starting to look, since it's at the perimeter. But most of the time
  companies are blacklisting the exfil destinations that I used in the
  past---not the new ones I'm using.

  All of this data comes out and enters into an analytics and
  evaluations stage [stage 6], where I, my team, or others evaluate
  it, make some decisions, and sort of re-task or redirect based on
  what we've found and learned.

  That's what my life is, and you will notice that I didn't talk at
  all about my tools or exploits, because I can change them up easily.

  \interviewer \question{Why don't the networks and companies see you
    doing this reconnaissance, collection, and exfiltration?}

  \operator First, they aren't looking. Everyone is very focused on
  the perimeter. But it's not where the treasure is for me. The
  treasure is on the inside; the treasure is the data inside the
  target. That's where I'm focused, not the perimeter other than
  initially and for minor things here and there. Once I've established
  persistent clandestine access to the network, I'm easily traversing
  the perimeter, in both directions as needed.

  Second, when they do look, my victims aren't thinking like an
  operator running a campaign. They're looking for point-in-time
  alerts, or a proverbial needle in a haystack. They're not looking
  across various data sources, they're not linking these things
  together to see what kind of campaign narrative comes out of it.

  \interviewer \question{OK, let's go through these different
    activities inside your targets in a bit more detail. Walk me
    through your day to day after you're already inside your target,
    from the perspective of the operator. Talk about the initial, and
    any ongoing, reconnaissance activities.}

  \operator I'm spending a significant amount of time doing
  reconnaissance [stage 3]. There's a lot for me to discover and
  figure out about this network: the hardware, the software, the
  users, the security posture, etc. All of this is in support of my
  goal, which is to find the large repositories of data that I
  want. I'm sort of always looking over my shoulder and both ways
  before I cross the proverbial street. Luckily most modern enterprise
  networks are super noisy, allowing me to be relatively blatant in
  how I look around and discover stuff.

  \interviewer \question{You said ``blatant'', but the industry talks
    about sophisticated threat actors using low-and-slow attacks. How
    often are you doing this? Could it be easily detected? }

  \operator Well that's the thing, I get to conduct this sort of
  behavior based on how I perceive the threat environment. Sometimes I
  need to go low and slow, other times I can be a bit more blatant. I
  don't need to do active reconnaissance too frequently, I just need
  to get a good map to work with that is relevant to finding, getting,
  and exfiltrating the various data stores. I'm looking for all sorts
  of things; probably 75\% of what I discover isn't necessarily of
  interest to me specifically, but the data goes back for others to
  look at and evaluate. Sort of over everything is that there's a cost
  associated with what I do and how I do it. For example, walking up
  and down a /24 with a noisy scan is quick and effective, but leaves
  more of a ``signature'' than if I scanned just small clusters of
  workstations across a few days or a few weeks---or more. Same deal
  with individual hosts: I can learn a lot at the network level, but
  often I need to get a bit deeper on the individual hosts to gain a
  better understanding of the what and the how. But there's an obvious
  risk to this. This is what most network defenders tend to forget:
  every decision I make is a trade-off that has been considered and
  decided, before I ever hit the enter key.

  \interviewer \question{You've done or are doing this reconnaissance,
    and I presume that you start to identify where the data lives and
    what types of protocols and credentials access it. What happens
    then?}

  \operator This is the collection/staging [stage 4], that I mentioned
  before. And to be clear, as I just talked about for reconnaissance:
  every decision I make about how I execute my op has some type of
  cost associated to me. So, let's say I find the database with the
  data that I care about, and it's 50GBs of data. The database is
  internal only, so I have to take that data and put it onto a host
  that has Internet access. Moving 50GBs around the network might seem
  blatant, but people are almost never looking at this east-west flow
  from internal system to internal system. As a result, I can
  generally move my data around inside the network with relative
  ease. It gets a bit more complicated in networks that are segmented
  and have internal controls like firewalls and enclaves and other
  such things, but these harder target environments are actually
  exceedingly rare.

  By this time, my risk profile is considerably lower. I can start to
  mimic legitimate users, for example. This may not seem important,
  but it means that I often don't have to exploit and/or even implant
  machines inside the network. I can use legitimate creds and/or
  legitimate access paths to interact with internal resources and
  extract the data that I care about. What this means from a practical
  standpoint is that all those end-point security tools and such that
  are looking for malware/implants and all sorts of bad stuff on the
  inside---\emph{I blow right past those.}

  \interviewer \question{So where does that leave the defenders?}

 \operator Well, at this point in my campaign I'm pretty well
 entrenched. I likely understand your network and security posture at
 least as well as you [the defenders] do. I've probably captured a
 bunch of legitimate creds, I might even be logged into your security
 interface watching you create and respond to tickets. But even with
 legitimate credentials that work, it's not as if the behavior itself
 is legitimate. Do Claude's credentials get me access to the database?
 Sure do!  Do I want to ensure I only access the database with
 Claude's credentials when Claude is in the office and at his
 computer? Well, maybe \ldots but that's a cost to me that I have to
 consider depending on the operation. So, whatever workstation I am
 using as my base of operations, from which I will interact with the
 database (using Claude's creds), has a behavioral aspect to it that
 is almost certainly anomalous compared to how data is accessed and
 moved around the company in support of legitimate business. And
 that's the real kicker: creds or no creds, day or night, or whatever
 other environmental variable is at play: it's how, when, which
 direction, and at what volume, and the frequency I interact with the
 data in the environment that is important---and how
 that looks and compares to legitimate data accesses, flows, and uses.

  \interviewer \question{Ok so you've done the recon, you've found and
    moved and staged the data; what's next? I assume you get the data
    out?}

  \operator Of course, [stage 5] exfiltration is key. I have to get
  the information out before we perform the in-depth analysis on it.

  \interviewer \question{Why don't you pre-filter and only take out
    the things you need? That would reduce the amount you have to
    exfiltrate, no?}

  \operator Actually, there are several reasons why I generally won't
  pre-filter the data in the compromised environment. First, maybe
  there's good information in the volumes of data that I've identified
  that someone on the back end will find valuable and that I don't
  know is interesting ahead of time. I don't want to leave data on the
  table; it's opportunistic. Second, if my scripts and tools are
  compromised---which happens from time to time---I don't want the
  victim to know what I was looking for. If I give that away it can
  point back to me or my teams, which I don't want. Not giving away
  what I find most valuable in their data prevents the organizations
  from isolating that type of information and making it more difficult
  for me to get in the future.

  \interviewer \question{Are you exfiltrating information constantly?}

  \operator For the larger throws, it's not always a constant stream of data,
  but it isn't a one-time thing either. I'm usually exfiltrating some
  pretty decent chunks of data aperiodically over the course of the
  compromise. Things like screenshots of workstations and keylogs are
  going to come out relatively frequently if I'm trying to watch and
  identify a particular person, although these are at a pretty low
  volume. I see people try to identify the C2 channels or these other
  small keylogger throws but the defenders are buried in false
  positives when they look for such small amounts of data. Things like
  massive database dumps generally come out at higher volume but with
  less frequency. This is where network defenders could figure out
  what I'm doing, but the current focus is misplaced, in my
  opinion. There is a lot of reliance on discovering known bad
  IPs/domains and then blacklisting them. And everyone celebrates when
  they discover that a particular IP is a C2 server for some implant
  in some network. But that external infrastructure, that I control,
  is incredibly fluid. Not only do I setup new malicious
  infrastructure to use, but I can also do clever things like take
  over a legitimate site and use it maliciously, or create fake
  but real looking domains. Or, use legitimate services like cloud or
  social media. Bottom line is that I can get around basically all the
  things you use to try to stop me from getting data in and
  out. What's far more relevant is the nature of the behavior itself:
  what does the internal to external data flow look like, at what
  volume and what day/time, and how does that compare to how that
  host---and all other internal hosts---talk to various external
  sites.

  \interviewer \question{Now that you've exfiltrated the data out,
    what happens?}

  \operator Well that's the consumption and processing of the data [stage 6],
  the part that no one on the operations team wants to talk about
  because it's not so glamorous, but the analysts love it. Sure, I
  exfiltrate all this data. But now someone needs to go through that
  data and derive meaning from it. And from my perspective, some of
  that meaning is critical because the information gleaned from the
  exfiltrated data will be used to create additional requirements for
  me, to do new things or keep doing the same things or to stop doing
  something.

  \interviewer \question{Are you always active in a target?}

  \operator No, not at all. There's dozens of factors that dictate how
  often I'm active inside a target network. Some of them are sexy and
  interesting, like, in the somewhat uncommon case where the network's
  security team is good so I want to minimize my chances of getting
  caught. Some of them are not so sexy, like, my team is mostly out
  sick this week, which means no one to catch beacons and no one to
  pull data. Or that I'm running a lot of operations simultaneously,
  and I can only light up a few at a time to give them the appropriate
  attention and care they need. It's also incredibly variable day over
  day, week over week.
  There is rarely as much pattern or automation as people imagine.
  This is important because a lot of network defenders
  use time and patterns to try to discover malicious behavior.

  I have to balance my risk of being detected against doing my
  job. Every time I'm interactive on the network there’s the chance
  that Murphy will pop up and something will go wrong and I risk being
  discovered, so I do what I need to do and then stop.

  \interviewer \question{What do you worry about?  What could your
    targets do that would make your job more difficult?}

  \operator Look at each of the stages from my perspective, identify
  activities that fit my work in each, and then connect across those
  stages. Understand how I do what I do, abstract the specifics away
  from the detection---like signatures and tools and exploits and
  known bad infrastructure---and instead focus on the behaviors within
  the network.
\end{interview}

\section{Fundamental Requirements}
\label{sec:requirements}

We believe this difficult problem---detecting
threat actors inside a network---is solvable.  There are four fundamental
requirements to achieve effective detection with a reasonable,
practical amount of resources:
\begin{enumerate}
  \item Focus on the threat actor campaign goals---not tools or artifacts
    of a single aspect (e.g., do not focus on zero day exploits,
    specific failed logins, etc.).
  \item Automation, machine learning, and interpretability.
  \item Adapt to an ever-changing environment.
  \item Hard for threat actor to avoid.
\end{enumerate}
Any approach that lacks these necessary pieces will not scale to large
networks or will lag behind evolving threat actors.

\subsection{Focus on Threat Actor Campaign and Data Movement}
\label{sec:mission-focus}

Focusing on the threat actor campaign, in a holistic way, is essential to
finding the threats that matter.  This is a monumental shift in focus
away from identifying and alerting on discrete anomalies.  Instead,
the solution must synthesize and make sense of activity across the
campaign lifecycle.

This is important because behaviors are most meaningful in the context
of the larger campaign.  For example, if a computer:
\begin{itemize}
  \item maps all the IP addresses in an environment,
  \item \emph{or} downloads 250 MB from an internal server it never
    communicated with before,
  \item \emph{or} uploads 15 MB to an external file server,
\end{itemize}
that could be mildly interesting.  Realistically, there is a high
chance each of these anomalies, individually, is benign.  But if the
same computer performed all three activities,
that is much more interesting.  It is \emph{especially}
interesting if this is not consistent with the computer's historical
activity.  The facts start to point more towards threat actor goals than
legitimate business functions: \emph{it is unlikely that all three
  anomalies co-occur for the same computer by random chance.}

\begin{figure}[htbp]
  \centering
  \includegraphics[width=0.95\textwidth]{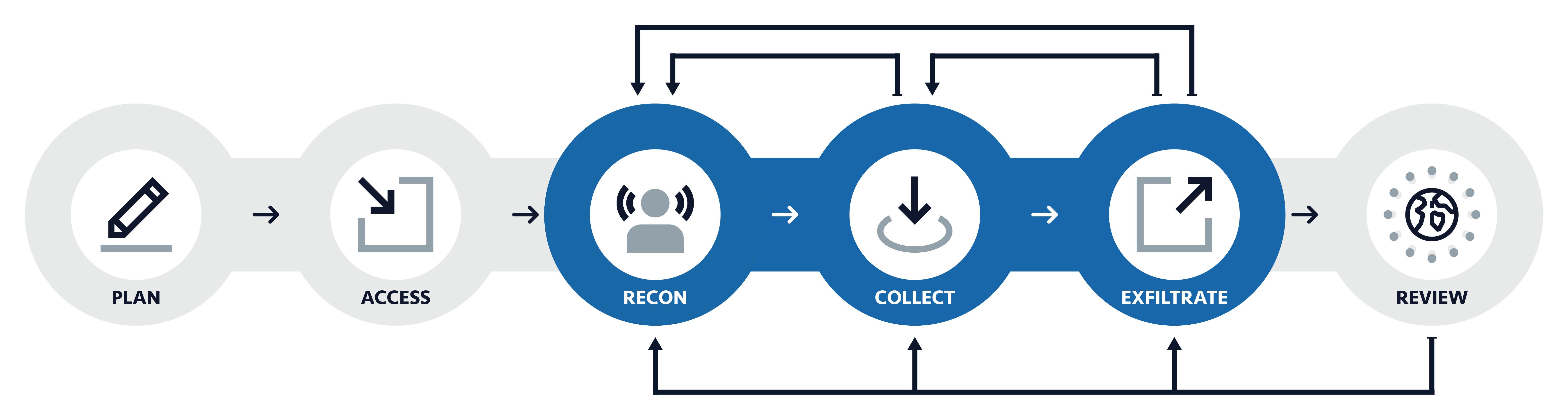}
  \caption{\small Campaign lifecycle for significant compromises.
    The blue stages occur inside the target network and therefore can be
    observed by network defenders.
    Recon, collection, and exfiltration will repeat during a long campaign,
    as the campaign proceeds through multiple cycles.
    Note also that the order of the stages is not fixed.
    E.g., after exfiltration, the threat actor may collect more data
    due to information learned in stage 6 (review).  See
    text for more details.}
  \label{fig:stages}
\end{figure}

Using a campaign-oriented framework for detection also simplifies what
needs to be monitored.  You only monitor for behaviors the threat actor
\emph{must} perform, the ones they cannot avoid, to succeed in their
mission.  Non-essential behaviors (e.g., beaconing) can be changed by
a committed threat actor---often at little or no cost---while core
behaviors cannot be easily changed.  Further, non-essential behaviors
are often a source of false positives.

There are several stages in adversary campaigns~\citep{Zatko18:MissionFocus}:
\begin{quote}
\begin{description}
  \item[Stage 1: Planning] The threat actor campaign starts with up
    front planning to figure out how to get access and the initial
    targeting.  This occurs outside the network, involving both public
    and private sources, and many of the tactics never need to touch
    the target network itself.  As a result, defenders usually cannot
    detect this activity.
  \item[Stage 2: Initial Access] Next, the threat actor gains initial
    access on the target network using any of a broad arsenal of tools
    (e.g., phishing, exploiting unpatched software on Internet facing
    infrastructure, zero day exploits, etc.).

    Perhaps surprisingly, monitoring for initial access helps very
    little with finding an active campaign in the network.  The set of
    tools is highly variable (with threat actors constantly finding new
    ways in), and typically threat actors remove evidence of the
    initial access.  Failed attempts potentially can be seen by
    defenders in logs, but this information is not really actionable for
    detecting if a threat actor later succeeded.
  \item[Stage 3: Reconnaissance] After penetrating the network, the
    malicious actor works to improve their understanding of the
    network: typical hardware and software configurations, locating
    key data stores, what endpoint protection is deployed, etc.
    Reconaissance activities include enumerating and scanning
    computers on the network, finding new treasure-troves of data,
    going through local logs, and identifying high-value targets.
    Legitimate users and systems do not exhibit the same type of
    uninformed searching and learning the environment.
  \item[Stage 4: Collection and Staging] During collection and
    staging, the threat actor gets access to the data
    and prepares to exfiltrate it.  Sometimes the
    data can be exfiltrated directly from the computer storing it, but
    other times it is collected and staged on a different
    machine.  There are many reasons why a threat actor might move the
    data before exfiltrating it, including:
    \begin{itemize}
      \item Move it to a machine that is always on so that the data
        can be exfiltrated at chosen times (when defenders are less
        likely to notice).
      \item Avoid drawing attention on a computer that administrators
        or users watch.
      \item Avoid any risk of crashing an essential server.  
    \end{itemize}
    The volume and/or breadth of these data movements usually stands
    out as atypical compared to data accesses triggered by business
    activities.
  \item[Stage 5: Data Exfiltration] Finally, the captured data is
    moved out of the network to infrastructure the threat actor controls.
    The particular exfiltration method(s) used are largely determined
    by the services and pathways permitted on the target network and
    by the organization's overall security posture.  Regardless of the
    method, the large-scale, aperiodic movement of data to an external network is
    a measureable activity that can be detected.

  \item[Stage 6: Review] After a threat actor exfiltrates data from a
    network, their next steps are to review this information and
    determine how to use it to further their goals.
    Threat actors do not leave a network after data exfiltration when
    new data of interest are discovered in review.  They will perform
    additional reconnaissance, collection, and/or exfiltration
    activity as part of that effort.
\end{description}
\end{quote}
See Zatko~\citep{Zatko18:MissionFocus} for a complete discussion of
the campaign stages and mission-focused threat
detection.\footnote{Compared to other frameworks for describing what
  threat actors do, this campaign life cycle places greater emphasis
  on discovering and moving data.  Lockheed Martin defined a
  high-level campaign life cyle, calling it a \term{cyber kill
    chain}~\citep{Hutchins11:CyberKillChain} because defenders could
  ``kill'' an intrusion by disrupting any of the phases in the
  campaign sequence.  The main focus is stopping the threat actor from
  getting established enough to take actions towards their objectives
  (a catch all bucket containing stages 3, 4, and 5 from
  Figure~\ref{fig:stages}).  The MITRE ATT\&CK\texttrademark\
  framework~\citep{Strom17:ATTACK} catalogues adversary tactics and
  techniques, with a heavy emphasis on how a threat actor gains
  initial access and establishes persistence in a network.  The
  Mandiant Attack Lifecycle Model~\citep{Mandiant13:APT1} is the
  framework most similar to the campaign life cycle used here.  It
  gives equal weight to how the threat actor infiltrates the network
  and establishes pervasive access (including internal
  reconnaissance); data collection and exfiltration are lumped
  together in the \term{Complete Mission} stage.}

Because stage 1 is almost impossible to observe, and because
threat actors use highly customized, variable, tactics for stage 2, we
believe these stages will not play a significant role in finding
active campaigns after an intrusion.  But during stages 3-5, threat actor
behavior is visible to network defenders and is sufficiently
predictable that we can build effective detection.

\subsection{Automation, Machine Learning, and Transparency}

The detection solution needs to be automated, with computer programs
doing the vast majority of the work.  To find campaigns, we need to
look at, detect, and cross-correlate relevant campaign behavior.
Further, the detection and analysis should be able to look over a period
of time to identify spread out campaign movements whose aggregate
volumes are large.
  Even with the savings of
only monitoring for the behaviors that are intrinsic and unavoidable,
there are still hundreds of outliers every day on a network.  It is
not practical for a human expert to comb through and assimilate weeks
of outliers to detect a campaign.  This is a machine-scale problem.

A long term solution will necessarily use machine learning to
discover, from data, how to define which activities are benign and
which are part of an adversary's campaign.  Within a network, what is considered normal and
appropriate depends on many things, like the role of the computer:
client, server, infrastructure, etc.
Further, what is normal varies across networks and over
time.  It is not practical or sustainable to manually write high
precision, nuanced rules to define ``normal'' in a program---the
program needs to be different for every network, and it likely needs
to be updated as the organization evolves.\footnote{The difficult part
  is writing specific rules (\emph{high precision}) that account for
  the network's context (\emph{nuanced}).  Simple, general rules can
  be written that define global outliers as abnormal.  These will
  catch some important campaign behaviors, but with the drawbacks of
  mislabeling some business-as-usual activities and not detecting more
  patient and less blatant campaign activities.}  Luckily, we can use
machine learning tools to learn what is normal from historical data.

We need to clarify what we mean by anomalies, how they relate to core
business functions on the network, and how finding them relates to the
campaign focus in Section~\ref{sec:mission-focus}.  General purpose
anomaly detection is probably not a sufficient solution to defining
normal vs. abnormal.  It finds too many benign anomalies, despite
decades of research going back to the
1980's~\citep{Denning87:IDES,Smaha88:Haystack,Javitz91:IDES,Porras97:EMERALD,Patcha07:ReviewAnomaly}.
Instead of looking for \emph{any} unusual activity, the solution needs
to be purpose driven: look for unusual activities that align with the
core campaign behaviors (Section~\ref{sec:mission-focus}).  In other
words, would the anomaly directly move an adversary campaign closer to
its mission goals?  An anomaly can either be an inappropriate action
(not part of the computer's core business function), or it can be a
legitimate action (part of the computer's core business function) but
taken to an extreme.

Abnormal activity is not automatically malicious, and \emph{normal
  activity is not automatically benign}.\footnote{For example, suppose
  the threat actor compromises the network administrator's computer and then
  logs into a router to change the network segmentation policy,
  unblocking access to a valuable data repository.  These are
  business-as-usual actions for both the user account and the computer
  involved, but they further the threat actor's campaign.}  This is why an
effective solution needs to \emph{automatically} correlate anomalies
and connect them---sometimes with normal behaviors that might have
served the campaign---so that they make sense and show the big picture
of a potential campaign.

\begin{figure}[htbp]
  \centering
  \includegraphics[width=0.90\textwidth]{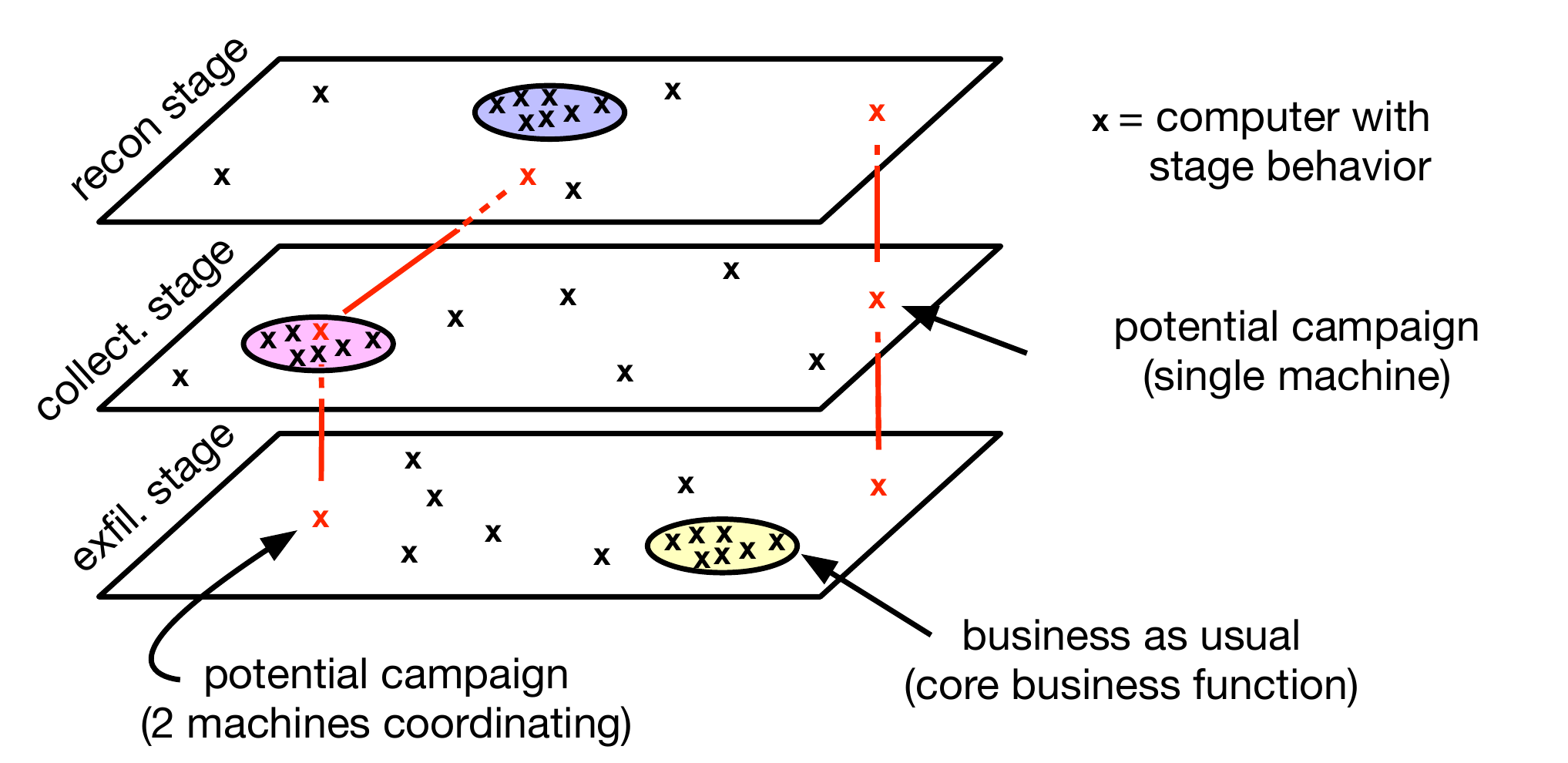}
  \caption{\small A threat actor's campaign is often a mix of normal and
    abnormal activities, and it stands out from benign anomalies if
    you connect the related risky behaviors.  We can view every
    computer on the network from three perspectives simultaneously:
    their reconnaissance-like, collection-like, and exfiltration-like
    behaviors.  The points mark computers performing recon-like, or
    collection-like, or exfil-like actions.  In each plane there is a
    cluster of computers whose core business function, their
    \emph{business-as-usual activities}, is to scan, or to collect, or
    to publish data externally (respectively).}
  \label{fig:bau-anomalies}
\end{figure}

Figure~\ref{fig:bau-anomalies} illustrates how the correlation across
campaign stages weeds out most of the benign anomalies, and creates an
opportunity to discover masked behaviors that are superficially
normal.  While it is common to detect scans at the network perimeter,
detect viruses, rely on data loss prevention tools, it is not common
to stitch the various pieces of information together.  If this is not
automated, human experts would need to manually correlate the
anomalies and piece together the clues---and this is not a scalable
approach.

Finally, the solution needs to be transparent so that security
analysts can understand what the detection system is doing and the information
it surfaces.  They need to understand what the big picture is, what
the individual activities are that make up the big picture, and why
the sets of activity should be considered unusual.  Further, automated
systems are rarely perfect.  Network defenders need to
be able to sanity check and verify any potential campaigns before
starting mitigation.  Verification is much simpler and faster if the
system clearly communicates the clues and how they fit together.

\subsection{Adapt to an Ever-Changing Environment}
\label{sec:adapt}

In addition to being automated, the solution must be designed to adapt
to changing network environments without grandfathering in pre-existing campaigns.  A
static solution will quickly become stale and ineffective.  This has
several implications:
\begin{itemize}
  \item The solution should support using a variety of raw data
    sources to get visibility into as much of the network as possible.
    This is necessary to adapt to changing networks. For example, a
    company that starts running servers on cloud infrastructure (in
    addition to running servers on premises) will want to extend
    monitoring to include unusual traffic for those servers.  This is
    straightforward if new data sources, like cloud flow logs, can be
    added to the system.

  \item The machine learning components must train on data in situ, to
    understand what is expected and consistent\footnote{Threat actors
      want to avoid creating constant opportunities for network
      defenders to identify them.  Remaining hidden is important.
      Consequently, internal business processes are largely more
      constant in periodicity, size, duration, etc.} for each
    individual network.
  \item Machine learning models must be retrained and updated
    frequently to stay relevant.  E.g., daily.  Otherwise the
    understanding of what is normal can become outdated as the network
    evolves in response to organizational requirements.
  \item Therefore, the training of models must be automatic and not
    require human intervention.  Nobody wants to babysit the system.
  \item Because the system is trained on data in situ, and because a
    threat actor may already be in the network, the machine learning models
    must not grandfather malicious activity as normal.
    This was a shortcoming of learning systems in the past.
\end{itemize}

\subsection{Hard for Threat Actor to Avoid}
\label{sec:hard-avoid}

It should be hard for a threat actor to avoid detection.

The system should support using multiple log data sources when they are
available.  Using multiple data sources makes it hard to hide.  The
more data sources across and within each campaign stage, the greater
the ability to detect insiders and persistent threats.  It also
significantly increases the cost for a threat actor to hide,
alter, or remove evidence of activities---particularly when they vary
per organization.  And threat actors have finite resources.

Similarly, the solution should prefer to use network and server logs
when possible.  It is harder to avoid leaving clues in network logs
and server logs than in client logs.  Network and server logs tend to
be stored on well-secured devices that are further from the perimeter.
Accessing these logs for deletion or modification is trickier.

The system, as a whole, must understand that the majority of
individual anomalies are benign and avoid presenting all of them to
security analysts.  Otherwise a threat actor can hide in the flood of
false alarms.

Finally, it should be hard (time consuming and/or expensive) for the
threat actor to influence what the machine learning models learn.  If the
threat actor can manipulate the analysis, they can create blind spots to
hide in.

\section{How \AppName\ Works}
\label{sec:approach}

\AppName\ focuses on your data, specifically
detecting discovery, access, and movement behaviors that fit the model
of a threat actor's campaign.

\subsection{Overview}

\term{\AppName} is a data processing and
analysis engine that detects potential adversary campaigns.  It
is a software-only product that consumes log data, from multiple data
sources, that is stored in a data lake~\citep{Versive17:DataLake}.
Every day, \AppAbbrv\ ingests
raw information from the data lake, updates activity profiles for each
host on the network, and updates machine learning models of normal
activity levels. The \term{activity profile} is a set of statistics
that summarize the host's activities over days or weeks.  The machine
learning models monitor the profiles for anomalies related to
reconnaissance, collection, or exfiltration.  \AppAbbrv\ surfaces
relationships in the data that match multiple stages
of the adversary campaign lifecycle.  To do this, it correlates
anomalies across campaign stages and across associated computers to
produce \threatcases.  Each \threatcase\ summarizes the potential
campaign and explains why the hosts and traffic are of interest and
add up to a risky narrative.

Before we describe \AppAbbrv\ in more detail, we first summarize how the
design meets the requirements from Section~\ref{sec:requirements}:
\begin{itemize}
  \item The system is built on top of existing log data sources, with a flexible
    data pipeline.  This makes it easy to incorporate new log sources,
    as network configuration changes (\textit{adaptable}).  It makes
    hiding in the network harder and a unique challenge for each
    site (\textit{hard to avoid}).
  \item Monitoring targets, that embody unavoidable adversary
    campaign behaviors, are derived from log data (\textit{mission focus}).
  \item Activities are monitored over long time windows because
    sophisticated campaigns unfold over weeks or months
    (\textit{mission focus}).  This increases the cost for the threat actor
    to avoid detection (\textit{hard to avoid}).
  \item Monitoring models are built and updated using machine learning
    every day.  They are learned from the organization's data so that
    they are customized to that network and represent what is normal
    for that network (\textit{automated, adaptable, hard to avoid}).
  \item The monitoring models take as inputs contextual information
    about the host and its activity so that the definition of normal
    is context-aware (\textit{automated}).  For example, \AppAbbrv\ tracks
    the size of typical data transfers to common external destinations
    (based on how most computers on the network interact with the
    destination).  This information adjusts the definition for what is
    a normal data transfer size---per destination.
  \item The system uses white box machine learning models so we can include
    \term{reason codes} in \threatcases\ that help explain why an
    activity is considered anomalous (\textit{automated}).
  \item Hierarchical models synthesize anomalous
    activities to find hosts exhibiting reconnaissance, collection, or
    data exfiltration behavior (\textit{mission focus, automated}).
    Few normal systems in an environment do all of these.
  \item \Threatcases\ are generated only when host(s) exhibit high risk
    for multiple campaign stages (\textit{mission focus, automated}).
  \item \Threatcases\ include the suspicious hosts, their high risk
    activities, what other computers were involved, and details on the
    activities (\textit{automated}).
\end{itemize}
Now let us examine each part of \AppAbbrv\ in more detail.

\subsection{From Log Data to Monitoring Targets}

The first step in monitoring a network is deciding what is worth
measuring.  \AppAbbrv\ derives monitoring targets from the log data that
embody core campaign behaviors.

A \term{monitoring target} is a column in a database table that
quantifies how much (or how little) each host performed an activity we
want to monitor.  The concept is easiest to understand by looking at
some examples (Table~\ref{table:targets}).  For example, \AppAbbrv\ measures
the neighborhood of how many internal destinations a host contacted
over a month-long time window.  This variable aligns with system
reconnaissance: small neighborhoods mean low recon risk, and large
neighborhoods mean higher recon risk.  \AppAbbrv\ measures all the monitoring
targets (variables) and assesses how unusual and risky each measured
value is (see \S~\ref{sec:monitor-models}).  We call the variables
\emph{targets} because of how they are used in the machine learning
models (\S~\ref{sec:monitor-models}).

\begin{table}[htbp]
  \centering
  \small
  \caption{Example Monitoring Targets}
  \label{table:targets}
  \begin{tabular}{p{3.5in}cc}
    \hline
    \textbf{Monitoring Target(s)} & \textbf{Log Source} & \textbf{Visibility Into} \\\hline
    How many distinct, internal IP addresses did the host touch over the last month?  (via ICMP, SNMP, TCP, UDP) & Flow & Recon (System)\\\hline
    How many distinct, internal IP addresses did the host touch over the last month via data$^a$, service$^b$, and general$^c$ ports? & Flow & Recon (System)\\\hline
    How many distinct, internal reverse DNS lookups (aka, PTR requests) did the host make over the last month? & DNS & Recon (System)\\\hline
    For internal addresses, how many distinct data$^a$, service$^b$, and general$^c$ ports did the host try to connect over (last month)? & Flow & Recon (Service)\\\hline
    How many total bytes did the host consume from every internal address over the last month? & Flow & Collection\\\hline
    How many total bytes did the host publish to every external address over the last month? & Proxy or Flow & Exfiltration\\\hline
    \multicolumn{3}{p{6in}}{\hspace{1cm}$a$: Data-related ports can be queried for data.  E.g., 21 (FTP), 80 (HTTP), and 156 (SQL).}\\
    \multicolumn{3}{p{6in}}{\hspace{1cm}$b$: Service ports are used by well-known services.  E.g., 6660-6670 (IRC).}\\
    \multicolumn{3}{p{6in}}{\hspace{1cm}$c$: General ports are everything else.}\\
  \end{tabular}
\end{table}

Said differently, monitoring targets are signals that can be used to
detect signs of a campaign using reconnaissance, collection, and exfiltration.
Unlike most network security approaches, we deliberately use a small
number of monitoring targets (less than 25).  Each monitoring target
should measure activities that embody fundamental campaign behaviors.

It is worthy clarifying the relationship between behaviors and
activities (and therefore, monitoring targets) in \AppAbbrv.  A
\term{behavior} is an abstraction, a category, that maps to high-level
step(s) in a threat actor's mission plan.  In contrast, an \term{activity} is
something that actually happened on the network.  It is concrete and
observable.  A behavior can be exhibited through different types of
activities.  For example, there are many ways to exfiltrate data,
including: upload files via HTTP/POST, and a DNS covert
channel~\citep{Piscitello16:DNSCovert}.  Each monitoring target
summarizes one kind of activity.  As a result, \AppAbbrv\ typically uses
multiple monitoring targets per campaign stage (behavior).  This is
especially true for reconnaissance because there are sub-behaviors
that reflect sub-goals:
\begin{itemize}
  \item System reconnaissance: discovering what devices are on the network.
  \item Service reconnaissance: discovering what services target
    computer(s) are running.
  \item Data reconnaissance: discovering what information is available
    in a data store, like the local file system or a SharePoint site.
\end{itemize}
and because there are multiple good tactics to use for those sub-goals.

Computing the monitoring target values is a large scale data
processing problem.  Log data can be huge for large organizations.
For example, one of our customers generates 12 TB of logs per day (600
GB compressed).
Suppose we wanted \AppAbbrv\ to analyze the last four weeks of activity and
compare it to the preceding four weeks of historical data (to get a
better baseline for activities).  That means ingesting, distilling,
correlating, and making sense of 56 days of data, or 672 TB for this
customer.

To handle this scale of data, \AppAbbrv\ is built on top of Apache
Spark~\citep{Zaharia16:Spark}
and runs on a distributed compute cluster.  The required data
transformations distribute well, allowing \AppAbbrv\ to scale horizontally.
To handle larger log data, just run \AppAbbrv\ on a larger cluster.  The
compute cluster can either run on hardware the customer owns (on prem)
or on cloud infrastructure (e.g., Amazon's Elastic Cloud Computing).
To access the log data, \AppAbbrv\ reads in raw log data from a data
lake in HDFS or in Amazon's cloud storage (S3).  Log files are expected to be
organized into date-partitioned folders.  E.g.,
\begin{verbatim}
    flow/20180520/00.txt.gz
    flow/20180520/01.txt.gz
    flow/20180521/00.txt.gz
    flow/20180521/01.txt.gz
    flow/20180521/02.txt.gz
    flow/20180521/03.txt.gz
\end{verbatim}

A multi-step data pipeline turns the log data into the desired
monitoring targets.  The log files are parsed to extract fields and
create a structured, tabular data set.  Next, fields are standardized,
to ensure things like consistent casing and timestamp formats.
Importantly, this includes mapping dynamic IP addresses to consistent,
stable machine names---a requirement for reliably tying together
activities that span weeks.\footnote{Mapping to stable machine names
  is especially important: for devices like laptops that connect both
  from home (over VPN) and from the office; for dealing with changes
  in DHCP lease names/IP; and for distinguishing between cloud
  instances with different roles but that get the same private IP
  address at different times (due to demand spinning up and down of
  instances).
}  After standardization,
data are transformed to get them ready for aggregation.  This includes
many things, but generally the purpose is to label the records with
various categories that are used to divide the data during
aggregation.  For example, flow records are labeled as traffic to
internal vs. external destinations and categorized into semantic port
groups.  (The port groups are defined based on their value to the
\emph{threat actor}: what data can be accessed; what network information
can be learned; or system liveness.  All of this is reconnaissance but
differing in intention, effort, and the value
returned to the threat actor.  See also Table~\ref{table:targets}.)

Aggregation happens in two steps.  The initial aggregation computes
activity statistics that summarize the latest day of records.  The
final aggregation combines the single day aggregates to get aggregate
statistics over a window of days.  With some care in the
implementation, the windowed aggregates can be computed in time that
is independent of the window duration.  These multi-day statistics,
computed for every device on the network, are the monitoring targets.

There are several benefits from aggregating, with little downside for
the use case of detecting campaigns.  Aggregation provides a smoothing
effect that gets rid of noise: instead of seeing every random spike as
an anomaly, \AppAbbrv\ detects aggregate outliers.  This makes it easier
to see the trends that matter.  Further, while low-and-slow
campaigns~\citep{Symantec11:APT} are hard to observe over an hour or a
day, over time the activities accumulate and become increasingly
obvious and anomalous.  And of course, the scalability benefits from
data reduction are massive.  The aggregated data are smaller and
easier to do advanced analysis on.  For the customer mentioned above,
the single day aggregates are 1\% the size of the raw logs (7 GB
vs. 600 GB, both compressed).
The main downside to aggregation is losing fine-grained information
about when activities happened.  Instead of pinpointing the minute
when a campaign action occurred, \AppAbbrv\ identifies the host(s) and the
day(s).  This is enough information to pivot to a SIEM or system of
record for the relevant raw logs,
and \AppAbbrv\ can generate the appropriate queries to drill into findings
and submit them to these systems.  A second drawback to this
aggregation strategy is a longer delay between when the anomaly
happens and when it is detected and shown to security analysts.  This
is because monitoring targets are updated on a daily cadence.  Given
our focus on campaigns that span weeks or months, a 24 hour delay seems
acceptable.

Finally, integrating new data sources is straightforward.  (See
\S~\ref{sec:hard-avoid} for why using multiple data sources is
valuable.)  Data
pipelines are specified as graphs of data flowing between reusable
transformation \term{blocks}.  The graphs are defined using a
domain-specific language.  The block-based data processing is a kind
of flow-based programming~\citep{FBP06}.  \AppAbbrv\ can be extended both by
defining new processing blocks and by defining new data pipelines.

\subsection{Using Machine Learning to Understand Normal}
\label{sec:monitor-models}

Once we have a monitoring target, we need to decide which of the
values are unusual and are worth highlighting as more likely to belong
to an adversary campaign than a benign business function.  Our
approach is to build a \term{monitoring model} using machine learning;
this model predicts the expected values for a monitoring
target.\footnote{This is why we call the statistics monitoring
  \term{targets}.  They are the target values, the goal values, for a
  model to predict.}  Then we compare the
predicted values with the actual values to identify surprising
activities.  If there is a large discrepancy between predicted and
actual values, that means the actual value is unexpected according to
the definition embodied in the model.  In order to make accurate
predictions that account for nuances of the host and its activity, the
models use context as inputs to reduce false positives.

A key benefit of this approach is that we can leverage supervised
machine learning to solve an anomaly detection problem.
\term{Supervised learning} simply means that the correct answer is
provided to the algorithm that builds the model.  Specifically, the
learning algorithm looks at labeled example records---with input
values and the correct output (target) value---and estimates a
function that maps from inputs to the output.  The estimated function
is the learned model that can be used for predicting target values.

Normally a lot of work is required to label data with the correct
answers, but in this application we get them for \emph{free}.  The
correct answer---what actually happened---is right there in the logs.
We just need to transform and aggregate the logs into the measurements
we care about monitoring.  Then, every day, \AppAbbrv\ analyzes the
previous day's monitoring target values to detect surprising values.
The models to do that are
learned using the monitoring target values from the day before that.

\subsubsection{Digression: What if the Labeled Data are Dirty?}
\label{sec:dirty-data}

When we learn these models, we must assume that the labels are
imperfect, aka \term{dirty}.  We expect to find benign, one-off
outliers in network data that cannot be predicted.  Even more
concerning, a threat actor may already be in the network, and some
labels reflect their campaign actions from the previous days, weeks,
or months.  The models of normal need to be constructed in a way that
avoids treating pre-existing, active campaigns as legitimate, normal
network behavior.

Our primary strategy for solving this challenge is to setup the
machine learning problems in a way that prevents the model from
memorizing the patterns of individual computers.  First, there is a
single shared model per monitoring target (vs. one model per computer
on the network); model learning gets low error by producing accurate
predictions for the whole network, not just individual machines.
Second, we avoid model inputs (context) that are unique identifiers
for machines because they would give the shared model enough degrees
of freedom to memorize each device separately.  For example, \AppAbbrv\ never
uses the name or IP address of the device as a model input.  Instead,
inputs are selected that might be useful for predicting the activity
of multiple machines.  Third, the automatic feature engineering in \AppAbbrv\
(see Appendix~\ref{sec:autolearn}) uses minimum sample sizes to ensure
each derived feature is relevant for predicting how a \emph{pool} of
machines behave.

Another way to think about this is that we always want to compare a
machine $M$'s activities to a group of computers.  Sometimes the
comparison group is the whole network (if none of the model inputs
provide useful context), and sometimes the group is the set of
computers that share some property with $M$ (see
\S~\ref{sec:context-models} for examples).

\subsubsection{Building Contextualized Monitoring Models}
\label{sec:context-models}

Because the goal is to predict a numeric value, like the expected
number of bytes collected, \AppAbbrv\ uses non-parametric regression
learning.  A \term{regression model} predicts a numeric value, as a
function of some inputs.  There are dozens of ways to learn simple or
complex regression models from data, including linear regression and
neural network algorithms.  \AppAbbrv\ uses a sophisticated non-parametric
learning algorithm that is fast, scalable, transparent, and able to
learn non-linear relationships.  It includes automated feature
engineering for finding informative representations of numeric and
string-valued inputs (context).  Appendix~\ref{sec:autolearn}
describes this learning algorithm in more detail.

\begin{figure}[htbp]
  \centering
  \includegraphics[width=0.8\textwidth]{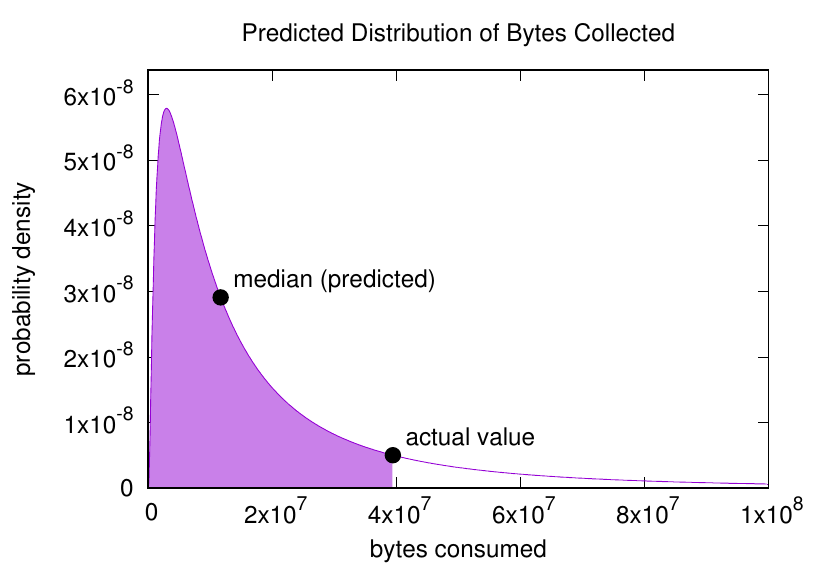}
  \caption{\small Actual values for a monitoring target are compared
    to the expected (aka, predicted) distribution to determine how
    surprising they are.  The further right the value lies in the
    distribution, the more surprising it is.  For example, the
    distribution above corresponds to the stage 4 collection model
    predicting 11.7MB collected, compared to 39.4MB actually
    collected.  The probability of seeing a more extreme value than
    39.4MB is 0.15 (the unshaded area of the distribution).  The
    location (median) and shape of the predicted distribution depend
    on the contextual inputs.}
  \label{fig:pred-vs-obs}
\end{figure}

After a monitoring model is learned, it is used to predict the target
values (aka, activity levels) for all devices on the network and score
their risk.  For example, the collection model predicts how many total
bytes each host consumed from every internal device during the last
month.  Each prediction sets the expected (normal) value, and this in
turn determines the probability of seeing a value as large as the
actual target value (see Figure~\ref{fig:pred-vs-obs}).  \AppAbbrv\ then
scales the probability to improve robustness and to better focus on
the activities subject matter experts consider most relevant to
detecting adversary campaigns.

\begin{figure}[tbhp]
  \centering
  \begin{subfigure}[b]{0.49\textwidth}
    \centering
    \includegraphics[scale=1]{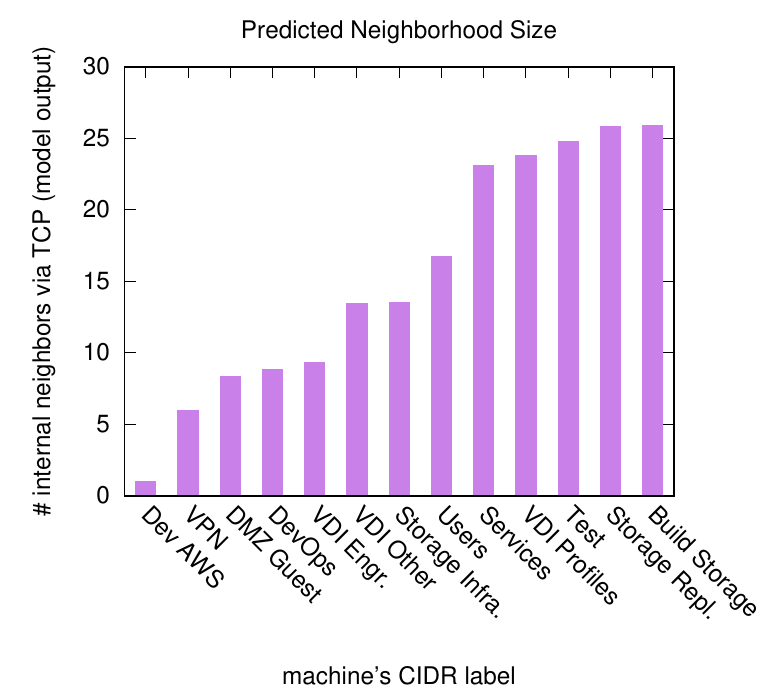}
    \caption{Stage 3 example.}
    \label{fig:pdp-cidr}
  \end{subfigure}
  \begin{subfigure}[b]{0.49\textwidth}
    \centering
    \includegraphics[scale=1]{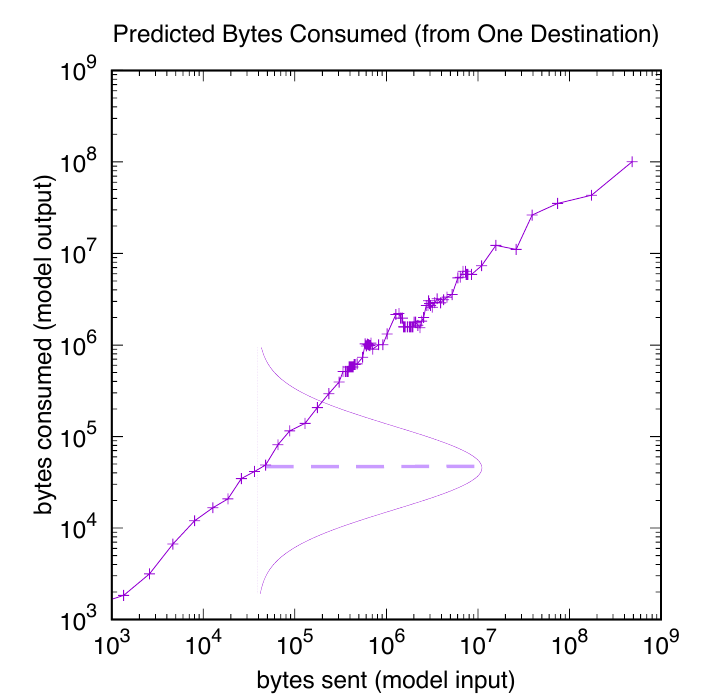}
    \caption{Stage 4 example.}
    \label{fig:pdp-bo}
  \end{subfigure}
  \caption{\small The expected activity (model output) changes based
    on context (model inputs).  These graphs show how the expected
    activity changes as a function of one particular input, after
    marginalizing over all other model inputs~\citep[p.~359]{HTF09:ESL}.
    (\subref{fig:pdp-cidr}) The expected neighborhood size ranges from
    1 neighbor to 20+, depending on the computer's CIDR block (network
    segment).
    (\subref{fig:pdp-bo}) The expected bytes collected by a computer
    varies greatly based on the number of bytes sent.  At 40 KB sent,
    the model predicts 50KB consumed.  Overlaid is the predicted
    distribution (see Figure~\ref{fig:pred-vs-obs}).  It would be
    suprising if the computer collected 10MB while sending 40KB in
    response.  In contrast, 10MB collected would be expected if 10MB
    had been sent.}
  \label{fig:context-changes-expected}
\end{figure}

Accounting for context, in the form of model inputs, is critical for
understanding what is normal for approved business
functions.\footnote{Recall the big picture here though.  A good
  understanding of network normal is helpful for detecting campaigns
  but not sufficient by itself.  In order to achieve effective
  campaign detection, pieces of evidence need to be linked across
  campaign stages (see \S~\ref{sec:mission-focus},
  Figure~\ref{fig:bau-anomalies}, and \S~\ref{sec:threatlinking}).
  Otherwise, you get standard anomaly detection
  results~\citep{Denning87:IDES,Smaha88:Haystack,Javitz91:IDES,Porras97:EMERALD,Patcha07:ReviewAnomaly}
  that are full of false
  positives~\citep{Axelsson00:BaseRateFallacy,Newman02:CryingWolf,Patcha07:ReviewAnomaly,Tavallaee10:CredibleEvaluation}.}
A model using context has a more nuanced and multi-faceted view of
business-as-usual.
Figure~\ref{fig:context-changes-expected} shows two examples resulting
from running \AppAbbrv\ internally on eSentire's network.

On the left, Figure~\ref{fig:pdp-cidr} shows how many internal
neighbors a computer $M$ is expected to talk to as a function of $M$'s
CIDR block (network segment).  At the low end, computers in
\texttt{Dev AWS} typically communicate with a single neighbor; this is
because our software development process involves engineers allocating
a development machine in EC2 and then tunneling into it.  Computers in
other CIDR blocks communicate with more neighbors, with the most
connected computers performing infrastructure and server roles (e.g.,
\texttt{Services, Test}, etc.).  On the right, Figure~\ref{fig:pdp-bo}
shows how the expected bytes collected by a computer $M$ from another
computer changes based on the traffic going in the other direction
(bytes sent).  This is because most network traffic involves two-way
communications.  \emph{In all cases, context can change the location
  and shape of the predicted distribution (recall
  Figure~\ref{fig:pred-vs-obs}), which leads to different actual
  values being considered surprising.}  One example distribution is
overlaid in Figure~\ref{fig:pdp-bo}.

Context is also a way to explain evidence to analysts reviewing a
case.  For each predicted value, \AppAbbrv\ generates \term{reason codes}
that explain which inputs most strongly influenced the prediction.
For example, Figure~\ref{fig:evidence} shows one piece of evidence
from testing \AppAbbrv\ internally with a simulated campaign.  The list of
reason codes enumerate the factors driving the prediction and tell
users some context supporting why the actual value was surprising.

\begin{figure}[thbp]
  \centering
  \fbox{\begin{minipage}[b]{0.95\textwidth}
      \ipaddr{10.42.28.17} collected $\underbrace{\text{3.20 GB}}_{\text{actual value}}$ from \ipaddr{10.42.24.15}.

      \vspace{1em}
      \AppAbbrv\ predicted $\underbrace{\text{10.10 MB}}_{\text{expected value}}$ because:
      \begin{enumerate}
        \item baseline prediction was 338.22 KB without context,
        \item 15.03 MB were sent to \ipaddr{10.42.24.15} [26.27x multiplier]
        \item \ipaddr{10.42.28.17} is in the Users CIDR block (\ipaddr{10.42.28.0/22}) [1.83x multiplier]
        \item collected from \ipaddr{10.42.24.*} [0.62x multiplier]
      \end{enumerate}
  \end{minipage}}
  \caption{\small Example evidence with reason codes.}
  \label{fig:evidence}

\end{figure}

Many kinds of contextual information are useful for understanding what
are expected business activities, including:
\begin{itemize}
  \item Attributes about the device, or its role in the network.
    E.g., CIDR block labels as in Figure~\ref{fig:pdp-cidr} (for
    segmented networks), or the security groups a cloud instance
    belongs to.  These
    help define \term{peer groups}, that is, sets of devices that
    should behave in similar ways
    due to similar functions (DB clusters, ETL systems, mail servers,
    clients, etc.).
  \item Symmetric byte flow volumes, as shown in Figure~\ref{fig:pdp-bo}.
  \item The historical activity for the device.  E.g., to predict the
    number of internal PTR requests (reverse DNS lookups) made over a
    month, look back at how many were made during the previous month.
    The historical time window and monitoring time window do not
    overlap.\footnote{Returning to the dirty data discussion from
      Section~\ref{sec:dirty-data}, care must be taken when using
      historical context to avoid grandfathering in threat actors that
      are already in the network.  In \AppAbbrv\, models only adjust
      predictions based on history if a) the device exhibited a
      regular day-to-day activity level during the historical time
      window \emph{and} b) the day-over-day fluctuations were
      relatively small. Otherwise, the device's historical activity
      level is ignored because its history is too irregular.  By
      requiring that the historical time series be regular /
      consistent, we avoid learning that scattered historical spikes
      are normal.

      For example, if the device communicated with the same number of
      neighbors every day (in the historical window), then \AppAbbrv\ would
      use the historical neighbor count to predict the number of
      neighbors for the current time window.  Similarly, if there is a
      repeating day of week activity cycle in the historical window,
      \AppAbbrv\ will treat the history as a consistent time series and use
      the historical activity to predict the current activity.

      Generally, threat actors try to reduce their visible presence on a
      network, and as a result they tend to exhibit sporadic, not
      continuous and steady, actions.  A threat actor could try to game
      this by repeating the same kinds of campaign actions on a
      regular schedule, but arguably this would raise their detection
      risk in other ways and raise their operational costs.
      (Repeating the same actions on a schedule can be error prone,
      and it would require investing extra time and resources.)}

  \item How other computers in the network tend to communicate with a
    common destination.  A \term{common destination} is an internal or
    external server (or cluster of servers with similar network
    addresses) to which many computers on the network talk. For
    example, if a company used CrashPlan for offsite backups,
    uploading large data volumes to \textit{code42.com} could be
    normal and expected---and that can be discovered as a pattern
    because there is a peer group of computers using that external
    service in similar ways.
\end{itemize}
Some context comes from joining in extra data sets, while other
context inputs need to be computed from the raw logs using the same
kinds of data pipelines that compute the monitoring targets.

\subsection{From Evidence to \Threatcases}
\label{sec:threatlinking}

At this point, we have monitoring models and can use them to score
activities, for every device.
The remaining work is to synthesize this information and build
\threatcases.

\begin{figure}[htbp]
  \centering
  \begin{subfigure}[b]{1.0\textwidth}
    \centering
    \includegraphics[scale=0.5]{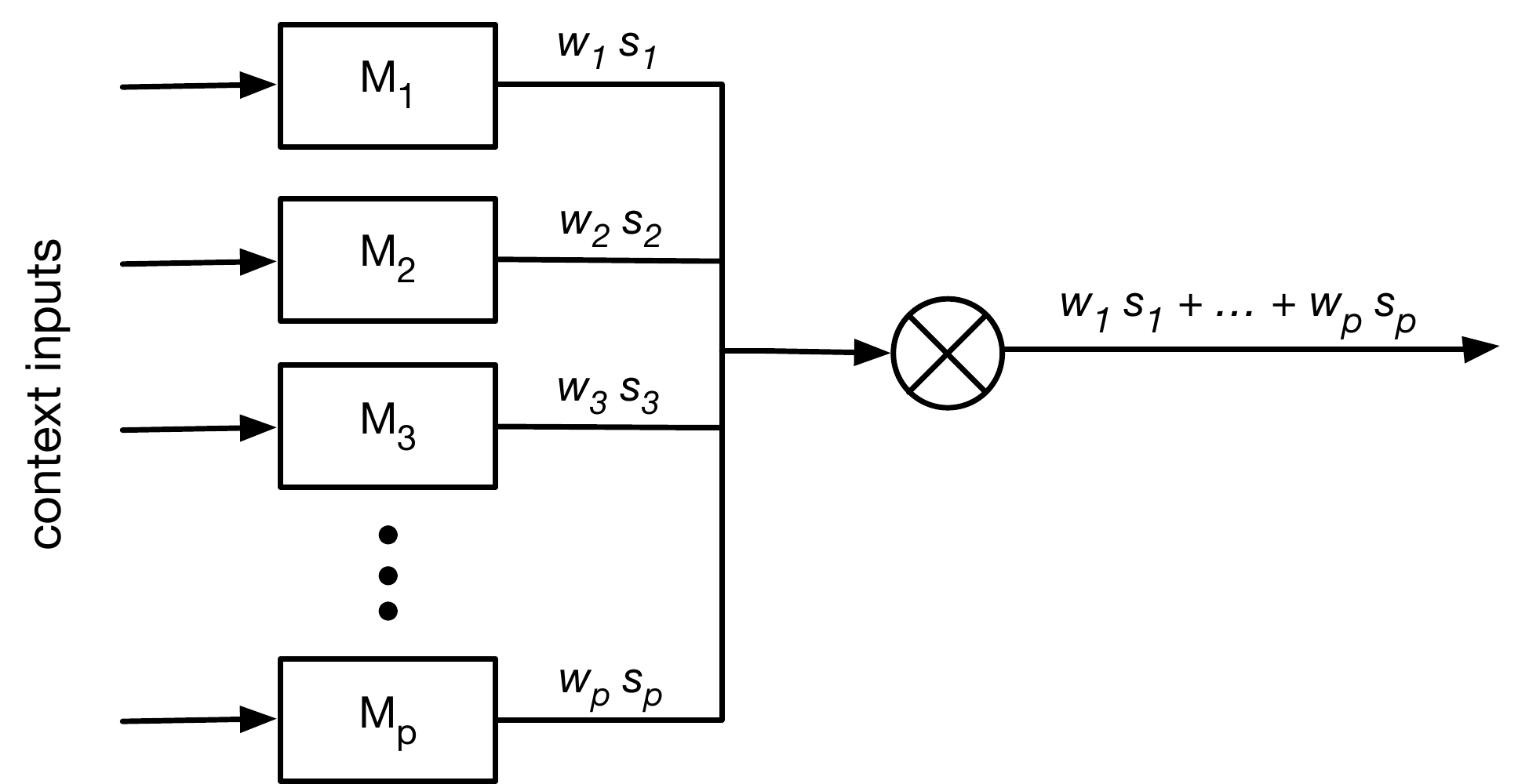}
    \caption{Reconnaissance ComboModel}
    \label{fig:recon-combo}
  \end{subfigure}\\
  \vspace{2ex}
  \begin{subfigure}[b]{1.0\textwidth}
    \centering
    \includegraphics[scale=0.5]{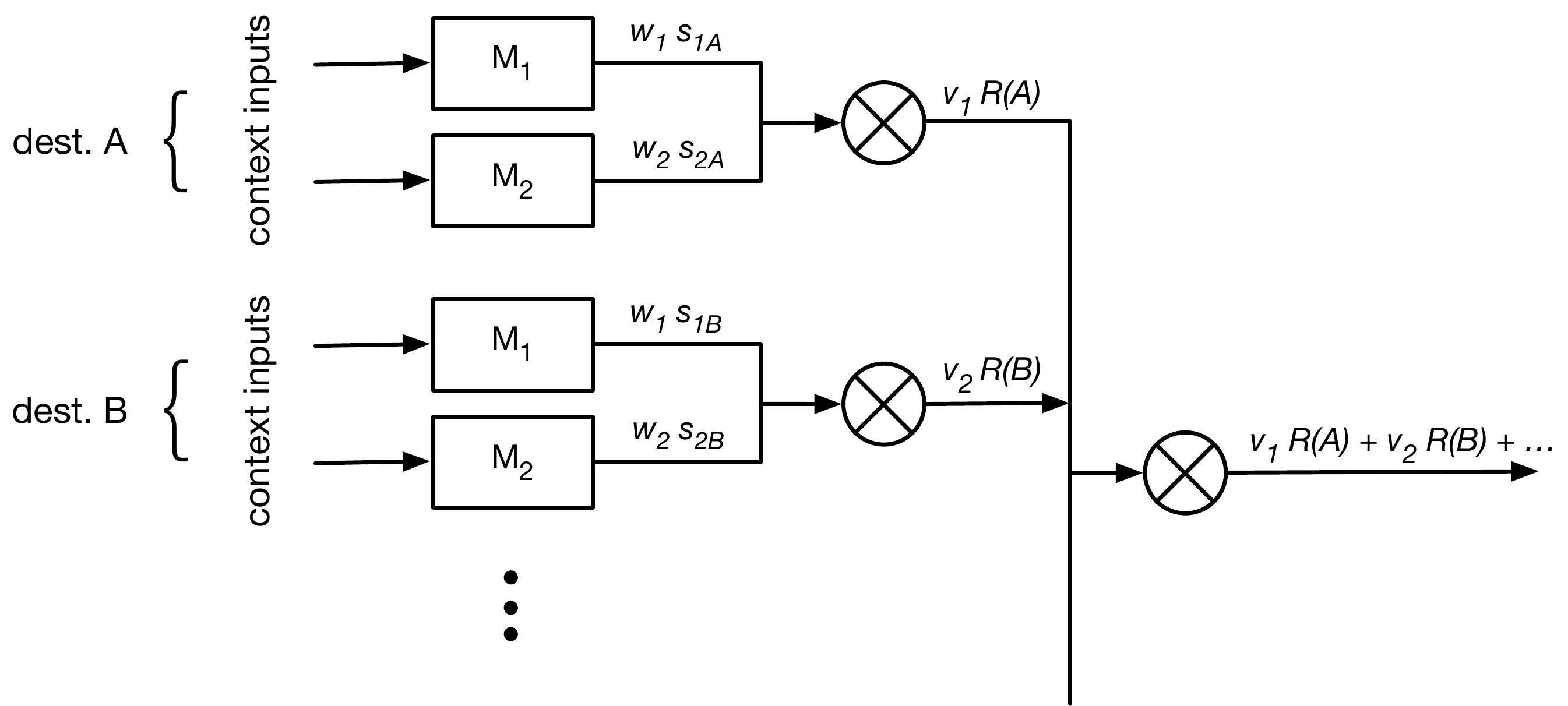}
    \caption{Collection or exfiltration ComboModel}
    \label{fig:granular-combo}
  \end{subfigure}
  \caption{\small ComboModels combine outputs from monitoring models
    into stage risk scores for a computer $X$.  The monitoring models
    $M_i$ output the surprise scores $s_i$.
    (\subref{fig:recon-combo})~The reconnaissance ComboModel performs
    a weighted addition of the surprise scores to get the total stage
    3 risk.
    (\subref{fig:granular-combo})~The collection and exfiltration
    ComboModels apply two levels of weighted addition to get the stage
    4 and 5 risk scores.  In the first level, monitoring models score
    the data flow for each destination (remote) device that $X$ talked
    to.  These are combined to get a risk $R()$ for the destination
    across all the monitored targets.  (Typically, there is one target
    per data source that captures flow information.)  The second level
    synthesizes the risk across all destinations, to get a total
    stage 4 or 5 risk for $X$.}
  \label{fig:stage-risk-model}
\end{figure}

\AppAbbrv\ first computes the stage-level risk scores for each individual
device to get an aggregated picture of campaign stage behaviors.  The
\term{stage risk scores} quantify the total reconnaissance,
collection, and exfiltration risk for each device, accumulated from
the scored activities in each stage.  For each stage, \AppAbbrv\ builds a
stage-level risk model, called a \term{ComboModel}, that synthesizes
the outputs of the relevant monitoring models.  The ComboModel is a
kind of \term{hierarchical mixture of experts} ensemble
model~\citep{Ruta00:fusion}.  The current ComboModels are simple by
design, using one or two layers of addition and expert-defined weights
to compose the monitoring models
(Figure~\ref{fig:stage-risk-model}).\footnote{Many architectures are
  possible for ComboModels, and it is not obvious a priori which
  one(s) are best.  We are actively exploring different architectures
  for the ComboModels to find structures that help highlight and tell
  the story of campaign behaviors.}

Recall that high risk for a single campaign stage is insufficient for
detecting potential campaigns with low false positives.  Network
devices often rank high on risk for a single stage.  \AppAbbrv\ should not be
surprised, for example, if a backup server scored high for data
collection.  This is normal behavior for a backup server, and it
should be ignored.  It is much less common for a machine to rank high
for two campaign stages, and it is exceedingly rare to score high for
all three.  When that happens, we have high confidence that the
machine is acting suspiciously and should be investigated.
Conversely, threat actors need to execute most or all of the recon,
collection, and exfiltration stages to complete their campaign mission
(\S~\ref{sec:mission-focus}), and it is costly for threat actors to hide
from detection in all campaign stages.

Consequently, \AppAbbrv\ integrates the stage-level rankings to get a single
ranking, with the highest risk devices at the top of the ranking.  \AppAbbrv\
uses the top-ranked devices as the starting points for building
candidate \threatcases\ (see below).  Internally, the total ranking
score is computed as the geometric mean of the device's stage-level
rankings.  Let $r_i$ be the rank position for campaign stage $i$.
Then:
\begin{align}
  \text{total ranking score} &= \mathrm{geomean}(r_3, r_4, r_5) = \left(\prod_{i=3}^5 r_i\right)^{\frac{1}{3}} = \sqrt[3]{r_3 r_4 r_5}
\end{align}
and ranking scores are sorted in ascending order.

We use the geometric mean because it is robust to a small number of
large values.  We want to ensure any devices with high risk in two
campaign stages are ranked high overall.  For example, if a
computer had the first, tenth, and thousandth highest stage risk
scores, its stage rank positions would be 1, 10, and 1000, and its
total ranking score would be 21.54.  In comparison, the normal
arithmetic mean of 1, 10, and 1000 is 337.
The single stage with low ranking (1000) dominates the
result and obscures the fact that the device ranked high in risk for
two stages (1 and 10).

At this point we have a ranked list of all the devices based on the
accumulated evidence of all their monitored activities.  You could
make simple \threatcases\ from this ranking, where each \threatcase\
consists of a single machine.  Frequently though, more than one
machine is involved.
Threat actors have become increasingly adept
in recent years at mounting campaigns that coordinate multiple
machines, with some used for recon, others for collection, and a third
group for exfil.  Therefore, when considering a high ranked device $M$,
\AppAbbrv\ also examines the risk profile of other machines with interesting
relationships to $M$.

\begin{figure}[htbp]
  \centering
  \includegraphics[width=0.95\textwidth]{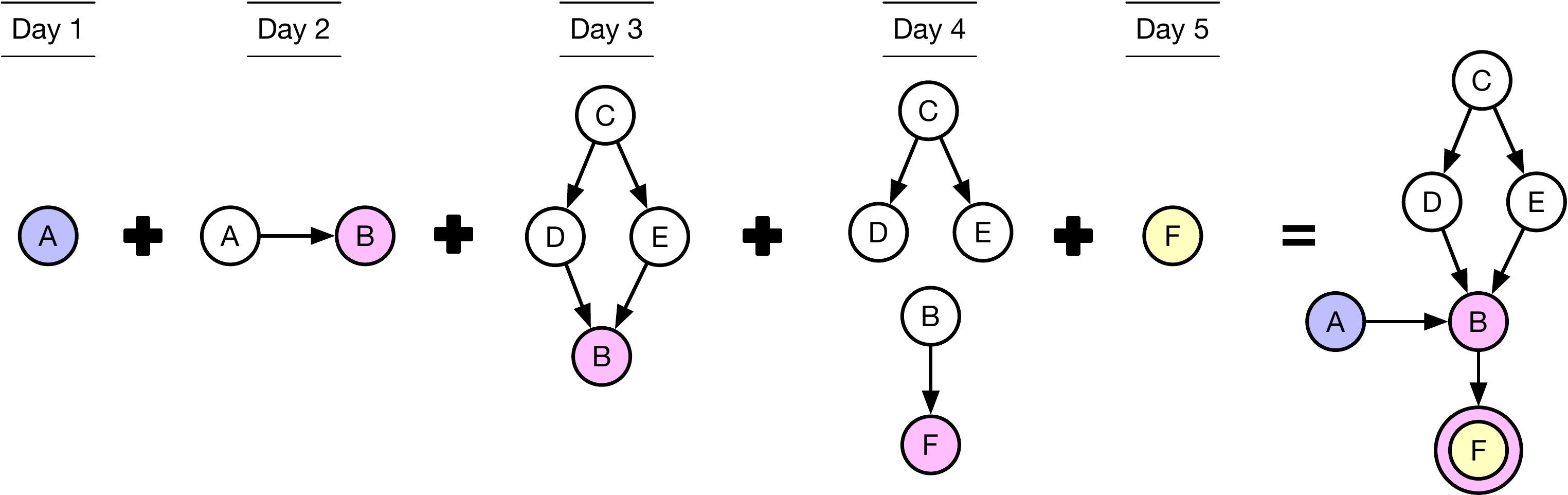}\\
  \vspace{1ex}
  \textbf{Anomalies}\\
  \vspace{1ex}
  \begin{tabular}{cccp{4in}}
    \hline
    \textbf{Day} & \textbf{Host} & \textbf{Stage} & \textbf{Activity} \\\hline
    1 & A & 3 & laptop $A$ contacts 73 machines on ports 22 (\textsc{ssh}), 110 (\textsc{pop3}), 139 (\textsc{smb}), 143 (\textsc{imap}), \& 445 (\textsc{smb}) \\
    2 & B & 4 & server $B$ consumes 50 MB from $A$\\
    3 & B & 4 & server $B$ consumes 600 MB from $D$ and 3 GB from $E$\\
    4 & F & 4 & HR workstation $F$ consumes 120 MB from $B$\\
    5 & F & 5 & uploads 120 MB to external server\\\hline
  \end{tabular}
  \caption{\small \AppAbbrv\ links together related suspicious computers
    to build holistic \threatcases.  This example shows five anomalies
    and the building of the related association graph (top right).
    Colored circles represent computers showing anomalous activity on
    the given day.  On days 3 and 4, there are large-but-expected data
    movements from $C$ to $D$ and $E$.  The final \threatcase\ would
    contain $A$, $B$, and $F$ as suspicious computers.}
  \label{fig:link-example}
\end{figure}

Multi-machine threat linking works as follows.  Starting from the
highest ranked computer, \AppAbbrv\ builds a candidate \threatcase\ where that
machine is the \term{seed}, or starting point, for the case.  \AppAbbrv\ then
adds other computers to the candidate \threatcase, where:
\begin{itemize}
  \item those computers received high or medium risk stage scores, \emph{and}
  \item the computers are immediate neighbors, reachable from the seed
    computer in the association graph.
\end{itemize}
The \term{association graph} contains one vertex per machine, and each
edge connects machines where there is an interesting relationship that
might indicate campaign coordination (e.g.,
Figure~\ref{fig:link-example}).  Like the monitoring targets, the
association graph summarizes activity over a window of time.  The
candidate \threatcase\ includes all the suspicous activities (as
determined by the monitoring models) for all the machines in the
\threatcase.  This process is repeated for the second, third, fourth,
etc. ranked computers, using each as a seed for generating a candidate
\threatcase.

The definition for association / coordination is flexible, and this is
one of the extension points for \AppAbbrv.  The default functionality is
to add a directed edge when there is a surprisingly large data
movement between the machines (i.e., the communication between them
gets high risk for data collection).  The edge points in the direction
of the data movement.

For example, in Figure~\ref{fig:link-example} there are three
suspicious computers to use as seeds for threat linking: $A, B,$ and
$F$.  These result in candidate \threatcases\ that contain $\{A, B,
  F\}$, $\{B, F\}$, and $\{F\}$, respectively.

The candidate \threatcases\ have high total risk---potentially
accumulated from risky behaviors on multiple machines---but they are
not guaranteed to include high risk for multiple campaign stages.
Therefore, in the final step \AppAbbrv\ filters the candidate \threatcases,
keeping those that contain high risk activities for at least two
campaign stages.

\section{Conclusion}
\label{sec:conclusion}

To detect persistent and insider threats, with a reasonable amount of
resources, there are four fundamental requirements:
\begin{itemize}
  \item Focus on the threat actor campaign goals.  Adopting a
    threat-actor-centric approach clarifies what the core campaign
    behaviors are, making it easier to see past the tremendous
    variations in tools and techniques.
  \item Automate as much as possible, including cross-correlating
    signals and behaviors, but also make the solution transparent to
    help security analysts separate persistent and insider threats
    from business-as-usual and acceptable noise.
  \item Adapt gracefully to an ever-changing environment and be robust
    to changes in threat actor tools and techniques to accomplish
    their campaign mission.
  \item The detection system needs to be hard for a threat actor to
    avoid.  This includes: using as many reliable, complementary data
    sources as possible; only showing analysts a small number of high
    confidence detections; and ensuring threat actors cannot easily
    influence what the machine learning models learn.
\end{itemize}

\AppName\ is designed from first principles to meet these
requirements.  Whereas other tools generate hundreds or thousands of
individual alerts per day with arbitrary weighting for risk,
\AppAbbrv's goal is to track multiple events across multiple adversary
campaign behaviors and surface just a handful of \threatcases\ per
week.  These cases should be high value because it is unlikely for a
correctly behaving system to exhibit multiple campaign behaviors by
accident.

To build a \threatcase, \AppAbbrv\ correlates anomalies across campaign
stages and links together associated, suspicious computers---without
requiring input from a security analyst.  The solution is dynamic and
adapted to the network being monitored: machine learning models are
learned from the network data (vs. pretrained and shipped with the
software), and are updated daily.  The \threatcase\ summarizes the
findings so that a security analyst can triage, verify, and launch a
deeper investigation quickly.

This approach is an important addition to a cybersecurity portfolio,
complementing prevention and rapid intrusion detection \& response.

\appendix
\newpage
\section{Automated Learning of Expressive, Transparent Models}
\label{sec:autolearn}

The eSentire learning library (esLL) takes labeled, tabular data and
returns a model that can predict what labels should be for future data
records.  The data table contains a target column (the labels, aka,
correct answer) and columns to use as model inputs.  The inputs can
be, and usually are, a mixture of numeric and string-valued columns.
The data table can be large and stored in a distributed data structure
in the memory of multiple computers in a compute cluster.

\subsection{Terminology}

This appendix uses the following terminology.

\begin{description}
  \item[Input] An input signal to a esLL model
    (Figure~\ref{fig:versive-model}).  A column from a data table that
    contains potentially useful information for predicting the
    modeling target.
  \item[Feature] A column in a matrix that is used as an input signal
    for a linear model.
  \item[Sparse Matrix] A data matrix that stores the non-zero values
    only.  When most of the values in a matrix are zero, using a sparse
    representation results in significant memory, and sometimes
    computational, savings.  For example, this matrix with three data
    records and five features:
    \[
    \begin{bmatrix}
      1 & 0 & 0 & 0 & 0.5\\
      0 & 0 & -2 & 0 & 0\\
      0 & -1 & 1 & 0 & 0
    \end{bmatrix}
    \]
    can be compactly represented by storing the indices and values of
    non-zero entries:
    \begin{quote}
      1:1 5:0.5 \\
      3:--2\\
      2:--1 3:1
    \end{quote}
\end{description}

\subsection{Background}

Before we can summarize the underlying learning algorithm, we need to
first review linear regression models and their limitations.  A
\term{linear regression model} is a function of the $p$ input
features $\vec{x} = x_1, x_2, \ldots, x_p$, where each feature is
multiplied by a weight coefficient $\beta_j$:
\begin{align}
  y &= f(\vec{x}) = \beta_0 + \beta_1 x_1 + \beta_2 x_2 + \cdots + \beta_p x_p = \beta_0 + \sum_{j=1}^p \beta_j x_j
\end{align}
For the simple case of $p = 1$, we end up with
\begin{align}
  y &= f(x_1) = \beta_0 + \beta_1 x_1
\end{align}
which is the equation for a line with slope $\beta_1$ and intercept
$\beta_0$.
These models are called \term{linear}
because the function is linear in the parameters: increase (or
decrease) $x_j$, and $y$ changes by a proportional, linear amount.

Linear models have much to recommend them.  There are multiple,
well-studied methods for estimating (aka, learning) the weight
coefficients $\vec{\beta}$ from data~\citep{Mandel82:SVDRegression,Malouf02:CompareMaxEnt,Minka03:CompareOptimizers,Nocedal06:NumericalOptimization,Hesterberg08:LARS},
including large scale data (large in terms of number of rows $n$ and
number of features
$p$)~\citep{Andrew07:OWL,Bottou10:LargeSGD,Crammer12:ConfidenceWeighted}.
For truly massive data, there are good algorithms for
distributing the learning computation across a compute cluster to
parallelize the work and achieve nearly ideal scaling~\citep{Chu06:MapReduce,Mann09:LargeScaleMaxEnt,Niu11:Hogwild}.
These models
are also wonderfully transparent: the model's equation can be printed
out and studied~\citep{Liao94:Interpreting}.
Further, individual predictions can be understood by
inspecting the input values $\vec{x}$ and the weights $\vec{\beta}$,
and looking for the largest factors (aka, terms) in the equation~\citep{Molnar18:InterpretableML}.

The main limitation of standard linear models is lack of
expressiveness: there are many functional relationships $y =
f(\vec{x})$ we would like to learn that are not linear. For example,
what if the input is a string and not a number: we cannot multiply a
string-valued $x_j$ by a weight coefficient.  What if the relationship
between $x_j$ and $y$ is quadratic, logarithmic, or a step function?
What if the true, unknown relationship is defined by two inputs
interacting with each other?  The assumption of a linear relationship
leads to inaccurate models in these kinds of situations.

However, there is a well known solution to extend the expressiveness
of linear models.  If we preprocess the data, we can change the
representation of the functional relationship into one that does have
a linear structure.  All of the examples above can be modeled well
with a combination of data preprocessing and linear regression (see
Table~\ref{table:preprocessing}).

\begin{table}[htbp]
  \centering
  \small
  \caption{Standard Preprocessing Tricks}
  \label{table:preprocessing}
  \begin{tabular}{lp{4in}}
    \hline
    \textsc{Scenario} & \textsc{Strategy} \\\hline
    String Input & Encode the string values as a set of binary indicator features, with one feature per string value.  For example, if the input is network protocol with possible values of TCP, UDP, and ICMP, then create three features where $(x_1, x_2, x_3) = (1, 0, 0)$ when the value is TCP, $(0, 1, 0)$ for UDP, and $(0, 0, 1)$ for ICMP.\\\hline
    Logarithmic Function & If the relationship between $y$ and $x_1$ is logarithmic, add feature $z_1 = \log(x_1)$ to get model $y = \beta_0 + \beta_1 x_1 + \beta_2 z_1$.\\\hline
    Polynomial Function & If the target is not a linear function of $x_1$ but instead a polynomial function, define appropriate features that are polynomial expansions of $x_1$.  For example, let $z_1 = x_1^2$, $z_2 = x_1^3$, and $z_3 = x_1^4$.  Then the expanded linear model would be $y = \beta_0 + \beta_1 x_1 + \beta_2 z_1 + \beta_3 z_2 + \beta_4 z_3 $.\\\hline
    Interacting Inputs & Add interaction features to the data set.  For example, the XOR function can be learned if we add feature $z = x_1 x_2$ to the data set and say the model's output is 1 when $y > 0$ and 0 otherwise.  Specifically, it can be solved by $y = 0.5 x_1 + 0.5 x_2 - 1.0 z_1$.\\
    \hline
  \end{tabular}
\end{table}

\subsection{The Learning Pipeline}
\label{sec:pipeline}

We are now ready to describe the learning algorithm in the esLL.  The
algorithm combines smart data preprocessing (sometimes called
\term{feature engineering}) with a core that learns generalized linear
models.\footnote{Generalized linear models is a class of model
  families that includes linear regression, logistic regression,
  multinomial classification, and Poisson regression.}  This
combination can learn non-linear models from large scale data
containing both numeric and string inputs.  Because the data
transformations are explicit and the core linear model is transparent,
these models can be inspected, and we can generate reason codes that
explain the factors that underlie any given prediction (e.g., Figure~\ref{fig:evidence}).  Importantly,
many decisions about data preprocessing can be decided automatically
by the esLL.

\begin{figure}[htbp]
  \centering
  \includegraphics[width=0.95\textwidth]{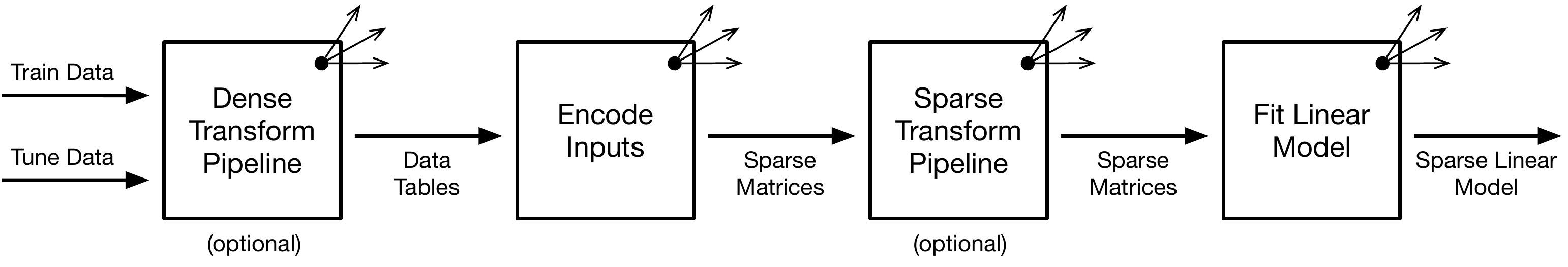}
  \caption{\small The learning pipeline in \AppAbbrv\ is a
    combination of smart data pre-processing and learning linear
    models.  Many variations are possible at each stage, resulting in
    many possible branches.  See sections~\ref{sec:pipeline}
    and~\ref{sec:exploration} for details.}
  \label{fig:learning-pipeline}
\end{figure}

Figure~\ref{fig:learning-pipeline} illustrates the steps in the
learning pipeline.  Beforehand, the labeled data are split into two
subsets, a training data set and a tuning data set.  The training data
are used to learn model weights, and the tuning data are used to
choose which model, out of many possible variations, is the best.

The \term{dense transform pipeline} transforms the data using a series
of table transforms.  These are operations that take a data table as
input and produce a new data table with new or modified columns.
Unlike the matrix transforms (see below), this part of the pipeline
has maximum flexibility for massaging the data.  For example,
\begin{itemize}
  \item String and timestamp columns can be manipulated directly,
    without special encoding.  For example, a transform could extract
    the day of week and whether that day was a holiday from a
    timestamp column, creating two new inputs for the model.
  \item Transforms can compute new columns based on multiple rows.
    For example, for input data with one row per day per computer, a
    rolling window transform can compute each computer's median bytes
    sent for the previous two months.
  \item Transform the target column to improve learnability.  In \AppAbbrv,
    for example, we add a replacement target column that is the
    $\log()$ of the original target column.  (The reason why this
    helps is technical and outside the scope of this paper.)
\end{itemize}
In our experience, the table transforms are usually chosen based on
subject matter expertise and domain knowledge.  Some transforms have
parameters that can be tuned automatically based on how it affects
model accuracy (e.g., how long should the window be in a rolling
window transform).

Next, the learning pipeline \term{encodes} the data table into a
sparse feature matrix.  The goal is to represent all the valuable
information in the data table as purely numeric values that are ready
for building linear models.  This is broken down by input column in
the table; for each input, an \term{encoder} function is applied to
produce a set of sparse features.  For example, a simple encoding
strategy for string inputs is to generate binary indicator features
with one feature per string value (see
Table~\ref{table:preprocessing}).  The snippets of features are pasted
together to get the full matrix.

Encoding is special in the learning pipeline because a) it always
happens, and b) it almost always needs to \emph{learn} the encoding
functions from data.  For the example of encoding a string input as a
set of indicator features, the esLL collects the set of possible values
from the train and tune data and then builds the encoding function
from that set.  As a second example, a useful encoding strategy for
numeric inputs is to sort the values and divide them into equal volume
bins, and then have one feature per bin that indicates if the input
value was in the bin's range.  The fence posts that set the bin
boundaries are function parameters that are learned from the data.
How many bins to use can be specified by the library caller, or it
can be learned from the data.

There are multiple choices for how to encode each type of input
column, and the best encoding strategy depends on the problem and can
be different for each input column.  The esLL includes the following
strategies:
\begin{description}
  \item[Passthrough Encoder (numeric)] Copy the numeric input values,
    unmodified, into a single feature column.
  \item[Quantile Binning Encoder (numeric)] Compute the quantile
    statistics (aka, fenceposts) for the input and create one binary
    feature for each of the discretized ranges (aka, bins).
  \item[Tree Encoder (numeric)] Learn a classification or regression
    tree~\citep{cartbook,Quinlan1993:C45} to predict the target, using only a single numeric
    input column.  The leaves in the tree divide the numeric values
    into discrete ranges; create one binary feature per tree leaf~\citep{Kohavi96:discretization}.
    This strategy is self-tuning: it learns both the number of bins
    and the location of bin fence posts automatically.\footnote{The
      tree encoder strategy is also interesting because it uses a
      standard non-linear machine learning algorithm, tree learning,
      as a subroutine to learn a feature representation for more
      expressive linear modeling.}
  \item[One-Hot Encoder (string)] See description in
    Table~\ref{table:preprocessing}.  This works well when the set of
    possible values is finite and relatively small; e.g., a set of
    department names or CIDR block labels.
  \item[Cluster String Encoder (string)] This is useful when the
    number of unique string values is large.  E.g., the input is the
    domain on the Internet where data was uploaded.  It is a
    simplified variation of \term{target
      statistics}~\citep{Micci-Barreca01:TargetStatistics}, where
    smoothing with a prior probability is avoided by requiring a
    minimum sample size.  Target statistics have been used in click
    prediction problems under the names counting features and
    historical
    counts~\citep{Chen09:BehavioralTargeting,Li10:EEAdvertising,Hillard11:QuerySegments,Ling17:ClickPrediction}.

    Cluster the data rows by the string values (one cluster per unique
    value) and compute the size and average target value for each
    cluster.  The averages represent the typical target value for data
    belonging to the cluster; intuitively, this should be very
    informative for predicting the target value.  The frequency
    (cluster size) summarizes how reliable the average is (more
    frequent observations equals more confidence that the sample
    average is representative of future observations).

    Next, apply the quantile binning encoder twice, once to the set of
    average values, and once to the frequency values, to generate
    indicator features that represent ranges of averages and ranges of
    frequencies.  Finally, generate interaction features between the
    average and frequency features.
\end{description}
\AppAbbrv\ uses all of these encoding strategies.

After encoding, the final preprocessing step is applying an optional
pipeline of matrix transforms.  A \term{matrix transform} operates on
a sparse matrix to produce a new sparse matrix.
The most important matrix transform is the interaction transform,
which derives new features by multiplying all pairs of existing
features together.  Because this includes multiplying a feature by
itself, the interaction transform implements both the polynomial and
interaction strategies from Table~\ref{table:preprocessing}.  The
interaction transform can be applied multiple times to get third,
fourth, etc., order polynomials and interactions.  E.g., $x_1^2 x_2$
would be a third order term generated after two applications of the
transform.

The last stage in the learning pipeline is fitting a linear model to
predict the target values from the given features.  The software uses
best practices for fitting robust models:
\begin{itemize}
  \item Use ridge~\citep{Marquardt75:ridge} or LASSO~\citep{Tibshirani96:lasso} regularization to control
    overfitting.  The regularization hyperparameter is chosen
    automatically by fitting models with different regularization
    strengths and keeping the variation with the best accuracy on the
    tuning data.
  \item Standardize each feature column in the matrix to have mean
    zero and standard deviation of 1.
\end{itemize}
The esLL uses the L-BFGS optimization
algorithm~\citep{Nocedal06:NumericalOptimization} when fitting the
model coefficients.

\begin{figure}[htbp]
  \centering
  \includegraphics[width=0.95\textwidth]{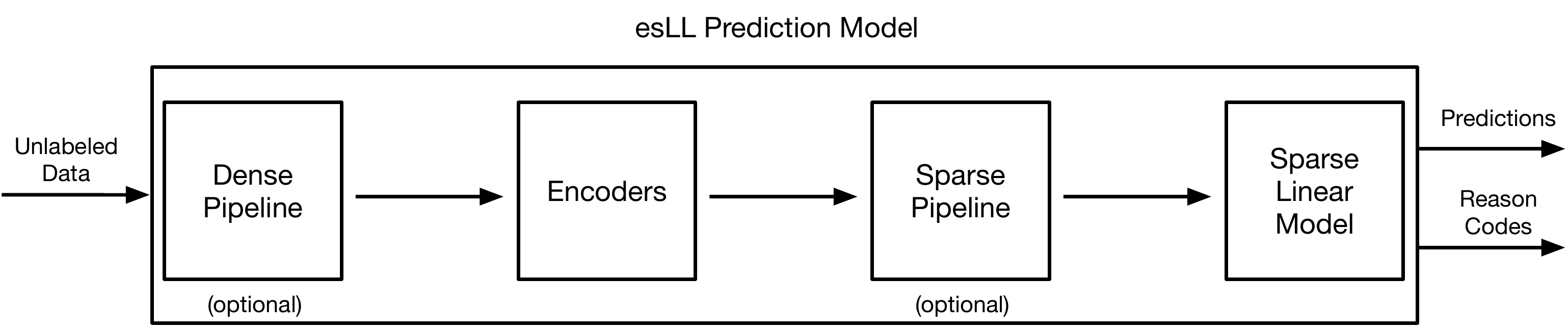}
  \caption{\small Anatomy of a esLL prediction model.}
  \label{fig:versive-model}
\end{figure}

The output of this learning pipeline is a prediction model that is a
non-linear function of the inputs (Figure~\ref{fig:versive-model}).
Internally, this model encapsulates all the transformations and
encoders that were used during the learning pipeline plus the linear
model that operates on the expanded feature space.  In contrast to the
learning pipeline (Figure~\ref{fig:learning-pipeline}), all of the
processing steps in the model are fixed functions with fixed parameter
values.  All choices about which encoders and transforms to use, how
to parameterize them, and the best model weight coefficients are
decided during learning.

The model internally stores the derivation for each feature.  This is
used to describe the model's logic and to generate reason codes
explaining how the top features influenced a given prediction.

\subsection{Automated Exploration}
\label{sec:exploration}

In the learning pipeline there are many
modeling decisions to make.  \textit{Which, if any, table transforms
  to use?  Which encoding strategies to use?  What hyperparameter
  values, e.g., number of bins, should be used? Should feature
  interactions be added during the sparse pipeline; second order,
  third order, etc.?  Which of the derived features should be kept
  (because they improve accuracy) and which should be discarded?}
This section describes how this learning pipeline is setup and
controlled to produce good models with minimal human effort.

Traditionally, a modeling expert and/or domain expert would spend
weeks experimenting with different pipeline variations, looking
for the combination that produced the best result.  Even with that
time investment, only a small fraction of the possible combinations
would be tried.  From a
usability standpoint, specifying all of the minute controls for the
learning pipeline would be tedious and error prone.

The esLL automates the trial-and-error search for a good model while
also allowing experts to control as much, or as little, of the
learning pipeline as they want.  To learn a model, the library is
called with a model blueprint, train data, and tune data.  The \term{model
  blueprint} is a configuration object that specifies the instructions
for defining the learning pipeline and building a model from the data.
Many instructions for a model can be left unspecified, and the esLL
will fill in the missing instructions using a combination of
heuristics and automated exploration to find good fill-in values.

\begin{figure}[htbp]
  \centering
  \includegraphics[width=0.8\textwidth]{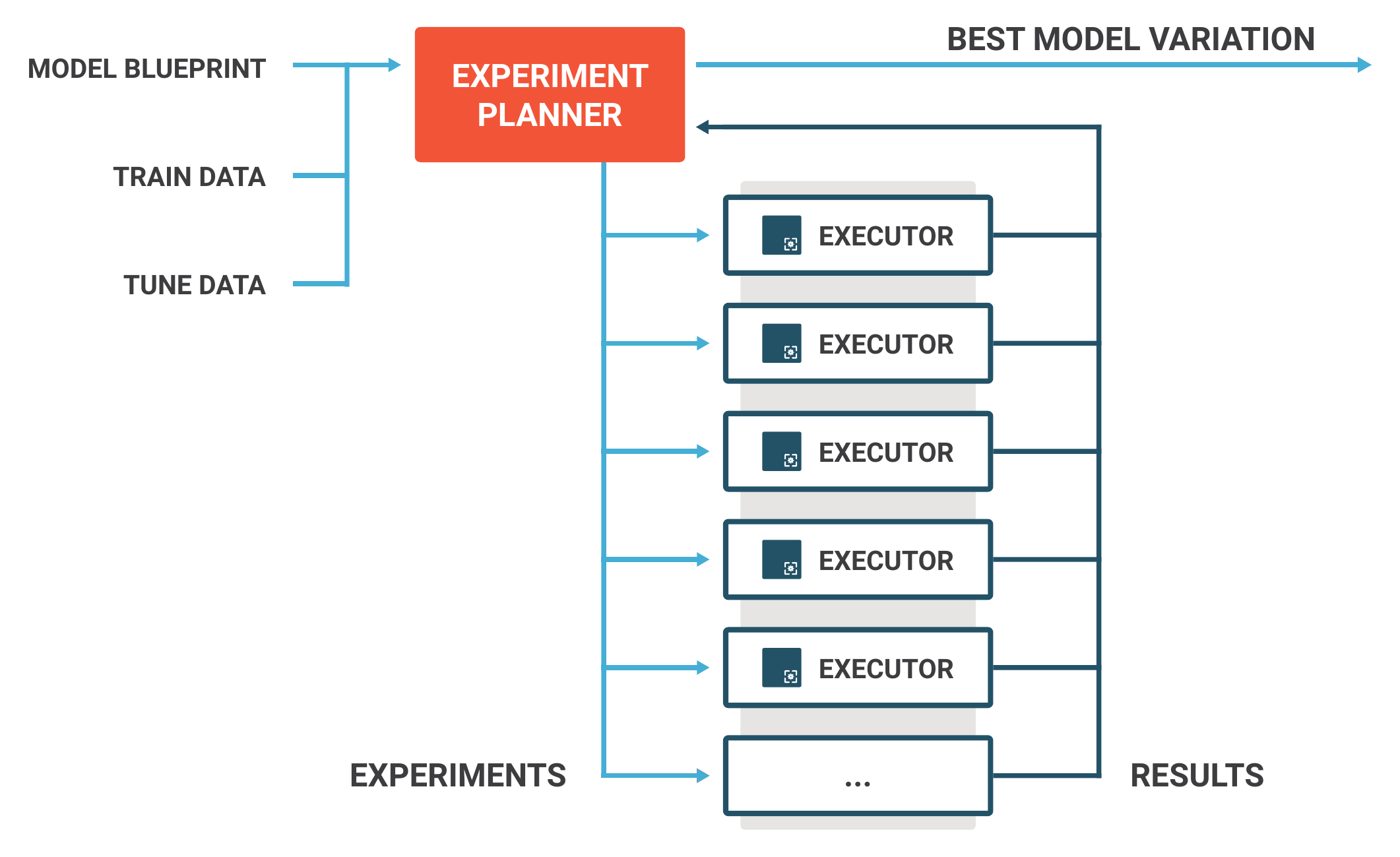}
  \caption{\small Flow chart for automated exploration.}
  \label{fig:auto-explore}
\end{figure}

Figure~\ref{fig:auto-explore} shows the control flow for this process.
An experiment planner takes the incomplete blueprint,
generates a set of experiments to run, and schedules them on a pool of
executors.  Each experiment is a full set of instructions for a
learning pipeline: which table transforms, which encoding strategies,
which matrix transforms, whether to use ridge or LASSO regularization,
and the generalized linear model family to use.  In other words, each
experiment asks the question of how good the model produced by the
particular learning pipeline will be.  An executor picks up an
experiment, runs the learning pipeline to produce a model, and then
evaluates the model's error rate on the tuning data.  The experiment results (the
error rates) are sent back to the planner.  The planner records the
best model found so far (and the instructions used to learn it). Based
on the results, it generates new experiments to explore new
refinements.  This continues until the best model is good enough, the
budgeted learning time has expired, every allowed model variation has
been tried, or progress stops being made.

Users control the extent of exploration with the
\code{time\_tradeoff} and \code{explorations} instructions in the
blueprint.  The \code{time\_tradeoff} instruction tells the planner how
much time should be spent exploring after learning a first
quick-and-dirty model.  For greater control, the \code{explorations}
instruction can be set to define the space of allowed explorations for
different part(s) of the learning pipeline.  For example, this is used
to instruct the experiment planner to try adding interaction features
during the sparse pipeline, and how many times to apply the
interaction transform.

In summary, the esLL learns quality models by automating
the brute force, trial-and-error work that used to be done by human
experts.  Thousands of variations are tried and
evaluated.  The net effect of automating
the learning pipeline is to automate formulaic feature engineering and
allow the human expert to focus on the domain and business problem.

\bibliographystyle{abbrvnat}
\newpage
\addcontentsline{toc}{section}{References}
\bibliography{references,conferences}

\end{document}